\documentclass{llncs}

\usepackage[title]{appendix}

\usepackage{amsmath,amsfonts}
\usepackage{algorithmicx}
\usepackage{graphicx} 

\usepackage{blindtext}
\usepackage{mathtools, cuted}
\usepackage{booktabs,chemformula}
\usepackage{textcomp}
\usepackage{blindtext}
\usepackage{subcaption}
\usepackage{lipsum}
\usepackage{booktabs}
\usepackage{siunitx}
\usepackage{adjustbox}
\usepackage[rightcaption]{sidecap}
\usepackage{url}
\usepackage{enumitem}
\usepackage{mathtools}
\graphicspath{ {figures} }
\usepackage[linesnumbered,ruled]{algorithm2e}
\usepackage[noend]{algpseudocode}
\usepackage{caption}
\usepackage{multirow}
\usepackage{array}
\usepackage{multicol, blindtext}

\usepackage[
colorlinks=true,
linkcolor=purple,
citecolor=purple,
urlcolor =purple,
]{hyperref}

\usepackage{flushend}

\usepackage{mathtools}

\newcommand{\comalgo}[1]{\texttt{\textcolor{red}{/\!/ #1}}}

\begin{document}

\title{Leveraging the Verifier's Dilemma to Double Spend in Bitcoin$^{\star}$}
 
%
%
\author{Tong Cao\inst{1,*}\and
 J\'er\'emie Decouchant\inst{2,\dagger}\and
 Jiangshan Yu\inst{3,\dagger}}
%
\institute{Kunyao Academy, Shanghai, China \and
Delft University of Technology, Delft, The Netherlands \and
Monash University, Melbourne, Australia}

\maketitle
\begingroup
\renewcommand\thefootnote{\inst{*}} 
\footnotetext{This work was partly performed while Tong Cao was with the University of Luxembourg.}
\renewcommand\thefootnote{\inst{\dagger}} 
\footnotetext{These authors are listed in alphabetical order and contributed equally.}
\renewcommand\thefootnote{\inst{\star}}
\footnotetext{This paper will appear on the 27th Financial Cryptography and Data Security conference (FC 2023).} 
\endgroup

\begin{abstract}

 
We describe and analyze \emph{perishing mining}, a novel block-withholding mining strategy that lures profit-driven miners away from doing useful work on the public chain by releasing block headers from a privately maintained chain.
We then introduce the \emph{dual private chain (DPC) attack}, where an adversary that aims at double spending increases its success rate by intermittently dedicating part of its hash power to \emph{perishing mining}. 
We detail the DPC attack's Markov decision process, evaluate its double spending success rate using Monte Carlo simulations. We show that the DPC attack lowers Bitcoin's security bound in the presence of profit-driven miners that do not wait to validate the transactions of a block before mining on it.

\keywords{Bitcoin \and Double spending \and Block withholding attack}

\end{abstract}




\section{Introduction}
\label{intro}




Bitcoin's security level is traditionally measured as the proportion of the mining power that an adversary must control to successfully attack it. Nakamoto assumed that an adversary would not control the majority of the mining power~\cite{nakamoto2008bitcoin}. If this assumption does not hold, an attacker is able to spend a coin twice and affect the system consistency in what is known as a double spending attack or 51\% attack. The soundness of the honest majority assumption has been discussed in the literature and mechanisms have been proposed to harden the mining process against the 51\% attack without completely eliminating it~\cite{Bonneau16,yu2019repucoin,HanSYL021,BadertscherLZ21}. 


Despite rewarding miners with newly minted coins and transaction fees, the Bitcoin mining process has also been shown to be vulnerable to selfish behaviors. Using selfish mining, a miner withholds mined blocks and releases them only after the honest miners have wasted computing resources mining alternative blocks. Selfish mining increases a miner's revenue beyond the fair share it would obtain by following the default Bitcoin mining protocol~\cite{selfish14}. Using simulations, selfish mining has been shown to be profitable only after a difficulty adjustment period in Bitcoin for any miner with more than 33\% of the global hash power~\cite{Gervais16,negy2020selfish}. Variants of selfish mining further optimize a miner's expected revenue~\cite{sapirshtein2015optimal}. 

Additionally, miners face the verifier's dilemma~\cite{luu2015demystifying,teutsch2019scalable,alharby2020data}, where upon receiving a block header they have to decide whether they should wait to have received and verified the corresponding transactions, or whether they should start mining right away based on the block header. Different miners might react differently to this dilemma.    


Following previous works, we say that a chain of blocks is public if the honest miners are able to receive all its content, while we say that a chain is private if some contents of the chain are kept hidden by the adversary.
In this paper, we show that an adversary can leverage a novel block withholding strategy, which we call perishing mining, to slow down the public chain in an unprecedented manner. More precisely, perishing mining leads miners that react differently to the verifier's dilemma to mine on different forks.
We then present the Dual Private Chain (DPC) attack, which further leverages the verifier's dilemma to double spend on Bitcoin.
This attack is, to the best of our knowledge, the first attack where an adversary temporarily sacrifices part of its hash power to later favor its double spending attack, and the first attack where an adversary simultaneously manages two private chains.
Intuitively, the first adversarial chain inhibits the public chain's growth, so that the second one benefits from more favorable conditions for a double spending attack.

To evaluate the impact of the distraction chain on the public chain we
first establish the Markov decision process (MDP) of perishing
mining. From this MDP, we obtain the probability for the system to be
in each state, and quantify the impact of perishing mining on the
public chain, i.e., its growth rate decrease. We further describe the
DPC attack and its associated MDP.
We then evaluate its expected success rate based on Monte Carlo simulations. 
Counterintuitively, our results show that the adversary increases its
double spending success rate by dedicating a fraction of its hash
power to slow the public chain down, instead of attacking it frontally
with all its hash power.

Overall, this work makes the following contributions.

$\bullet$ We present perishing mining, a mining strategy that is tailored to slow down the progress of the public chain by leveraging the verifier's dilemma. Using perishing mining an adversary releases the headers of blocks that extend the public chain so that some honest miners mine on them while some honest miners keep mining on the public chain, which effectively divides the honest miners hash power. 
We present the pseudocode of the perishing mining strategy, establish its Markov chain model and quantify its impact on the public chain growth.

$\bullet$ Building on perishing mining, we describe the DPC attack that an adversary can employ to double spend by maintaining up to two private chains. The first chain leverages the perishing mining strategy to slow down the public chain's growth and ease the task of the second chain, which aims at double spending. We provide the pseudocode of the attack, and characterize the states and transitions of its Markov chain model. 

$\bullet$ We evaluate the perishing mining strategy and the DPC attack based on extensive Monte Carlo simulations. Our results indicate that perishing mining reduces the public chain progress by 69\% when the adversary owns 20\% of the global power and 50\% of the hash power belongs to miners that mine on block headers without verifying their transactions. In comparison, selfish mining, which aims at optimizing a miner's revenue share, would only decrease it down by 15\%. 
Our evaluation also shows that an adversary that owns 30\% of the global hash power can double spend with 100\% success rate when 50\% of the hash power belongs to optimistic miners who do not verify transactions (i.e., type 2 miners in \S\ref{mtypes}).
While we focus on the double spending threat, we also show that the DPC attack allows an adversary to obtain a higher revenue than the one it would obtain by mining honestly or following previously known strategies (Appx.~\ref{profit}).

This paper is organized as follows. \S\ref{sec:background} discusses the related work and provides some necessary background. \S\ref{sec:models} defines our system model. \S\ref{sec:overview} provides an overview of the DPC attack. \S\ref{sec:dpc_spv} details the perishing mining strategy and the DPC attack that builds on it. \S\ref{sec:perf_eval} presents our evaluation results. \S\ref{sec:discussion} provides a discussion on other aspects of the attack. Finally, \S\ref{sec:conclusion} concludes this paper.

\section{Related Work} 
\label{sec:background}




\textbf{Double spending attack.}
The double spending attack on Bitcoin was described in Nakamoto's whitepaper~\cite{nakamoto2008bitcoin}, and has been further analyzed since~\cite{karame2012doublespending,rosenfeld2014analysis}.  Nowadays, $z=6$ blocks need to be appended after a block for its transactions to be considered permanent. 
An adversary with more than 50\% of the global mining power is able to use a coin in a first validated transaction and, later on, in a second conflicting transaction.
Nakamoto characterized the race between the attacker and the honest miners as a random walk, and calculated the probability for an attacker to catch up with the public chain after $z$ blocks have been appended after its initial transaction. 
Our DPC attack aims at double spending, and improves upon the classical double spending's success rate. 

\textbf{Block-withholding attacks.} Selfish mining was the first mining strategy that allows a rational miner to increase its revenue share~\cite{selfish14}, and was later shown to harm the mining fairness~\cite{bonneau2015sok,CromanDEGJKMSSS16}.
Selfish mining is not more profitable than honest mining when the mining difficulty remains constant despite the fact that the adversary is able to increase its revenue share~\cite{gobel2016bitcoin,Gervais16}. Nayak et al. proposed plausible values for the selfish miner's propagation factor by utilizing the public overlay network data~\cite{stubborn16}. They pointed out that the attacker could optimize its revenue and win more blocks by eclipsing~\cite{heilman2015eclipse} honest miners when the propagation factor increases. Gervais et al. analyzed the impact of stale rate on selfish mining attack~\cite{Gervais16}. Negy et al. pointed out that applying selfish mining in Bitcoin is profitable after at least one difficulty adjustment period (i.e., after approximately two weeks at least)~\cite{negy2020selfish}.
The DPC attack differs from these works in the sense that its main goal is not to increase the adversary's mining share but to double spend with higher probability than previous attacks.

\textbf{Combining selfish mining and double spending.}
Previous works have shown that an adversary can combine the double spending attack with selfish mining~\cite{Gervais16,sompolinsky2016bitcoin}. 
In this attack, the attacker maintains a single chain, which lowers the double spending success rate compared to the initial double spending attack.
Our DPC attack shows that an adversary can simultaneously manage two private chains to launch a more powerful double spending attack. 

\textbf{Blockchain Denial of Service Attacks.} 
The BDOS attack proposed strategies to partially or completely shut down the mining network~\cite{mirkin2020bdos}. To do so, the adversary only sends the block header to the network whenever she discovers a block that is ahead of the public chain and there is no fork, and publishes the block body if the next block is generated by the honest miners. By doing so, the profitability and utility of the rational miners and Simplified Payment Verification (SPV) miners is decreased, so that they eventually leave the mining network. The objective of BDOS attacks is to halt the system.
An adversary would need to spend approximately 1 million USD per day to shut down the system. 
Our DPC attack frequently separates other miners' hash power, which has some similarities with the BDOS attack's partial shut down case. However, the DPC attack allows the adversary to double spend.

\section{System Model}
\label{sec:models}


This section introduces the categories of miners we consider, and the adversary that launches a DPC attack. Table~\ref{table:1} summarizes our notations. 

\begin{table}[t]
	\centering
	\caption{Notations.}
	\begin{tabular}{|c|p{10cm}|} 
		\hline
		Symbol & Interpretation \\ 
		\hline
		$\alpha \in [0,0.5]$ & Mining power of the adversary. \\ 
		$\beta \in [0,1]$ & Fraction of its mining power that the adversary dedicates to its first private chain.  \\
		$\mu \in [0,0.5]$ & Mining power of type 2 miners. \\
		\hline
		$v_t$ & Value of the transaction the adversary inserts in a block when starting the DPC attack and attempts to double spend.\\
		$v_b$ & Mining reward per block.\\
		\hline
	\end{tabular}
	
	\label{table:1}
\end{table}

\subsection{Bitcoin mining and the verifier's dilemma}

Bitcoin mining is a trial-and-error process\footnote{https://en.bitcoin.it/wiki/Block\_hashing\_algorithm}. 
The public blockchain (or chain) is visible to all participants, and is maintained by honest miners. 
To achieve consistency, honest miners accept the longest chain in case of visible forks~\cite{pass2017analysis,gavzi2020tight,dembo2020everything}.
However, temporary block withholding attacks have been shown to threaten Bitcoin's security~\cite{karame2012doublespending,rosenfeld2014analysis,selfish14,sapirshtein2015optimal}.  Honest miners monitor the network to verify block headers and verify transactions. 

In the Bitcoin's network, block headers are often propagated faster than transactions. Bitcoin's incentive mechanism does not directly reward the verification of transactions, and BIP-152\footnote{https://github.com/bitcoin/bips/blob/master/bip-0152.mediawiki} introduced the compact block propagation optimization where each node can relay a block in a compact format before verifying its transactions. In this case, a miner that immediately mines on the block header of a correct block gets a time advantage to find the next block. If the miners instead wait and verify the included transactions before the next mining round, then they might sacrifice some non-negligible time in the mining race~\cite{luu2015demystifying,teutsch2019scalable,alharby2020data,cao2021characterizing}.

We assume that miners follow the traditional block exchange pattern~\cite{DeckerW13,mirkin2020bdos} using the overlay network. 
Block dissemination over the overlay network takes seconds, whereas the average mining interval is 10 minutes. While accidental forks (which may
occur every 60 blocks~\cite{DeckerW13} on average) reduce the effective honest mining power on the public chain and makes our attack easier, we do not consider accidental forks created by honest miners in order for simplicity. We evaluate mining and double spending strategies using event-based simulations where an event is the discovery of a block by a category of miner. 
We note $v_b$ the mining reward that miners obtain whenever a block they have discovered is permanently included in the blockchain.  

\subsection{Miner Categories}\label{mtypes}

We consider two types of honest Bitcoin miners that react differently to the verifier's dilemma: \emph{type 1 honest miners} and \emph{type 2 honest miners}.

\emph{Type 1 honest miners} always follow the default mining protocol and mine on the longest chain of fully verified blocks. In particular, these miners do not mine on a block header that extends a longer non-fully verified concurrent chain.

\emph{Type 2 honest miners} are profit-driven. As Bitcoin allows miners to accept and generate new blocks without verifying their transactions, type 2 miners start mining on a new block or its header if it extends the longest chain without verifying the transactions it contains. Note that type 2 miners can verify transactions whenever they are received and stop mining on a block header when associated transactions are faulty, or if they successfully mine the next block without having received the previous transactions.
In our experiments, we consider two opposite categories of type 2 miners that behave differently upon reception of successive block headers to evaluate the best and worst possible attack results.
\begin{itemize}[noitemsep,topsep=0pt]
\item \emph{Optimistic type 2 miners} miners always mine on the longest chain of blocks, which is possibly made of several block whose transactions have not yet been received. In particular, Simplified Payment Verification (SPV) miners~\cite{f2pool,halfspv,spvpools,btcstack} can be categorized as optimistic type 2 miners. Upon finding a block, optimistic type 2 miners can create an empty block or include transactions that they know cannot create conflicts (e.g., internal transactions for mining pools).
\item \emph{Pessimistic type 2 miners} only accept to mine on a block header if it extends a chain of full blocks. In particular, a pessimistic type 2 miner that extends a block header would then mine on the last block with transactions not to waste time. If the missing transactions eventually arrive, they then release the next full block. While if they extend over the last full block, they then create a fork.
\end{itemize}

In practice, it would be difficult for the adversary to identify the exact proportion of the global mining power that each type 2 miner subcategory represents. However, the adversary can be conservative and assume that all type 2 miners are pessimistic, since our attack still improves over the state-of-the-art in that case. We also discuss evidence for SPV mining in \S\ref{sec:discussion}, which is arguably the simplest type 2 mining strategy. 


\emph{The adversary} owns a fraction $\alpha \in [0,0.5]$ of the global hash power and its aim is to double spend with higher probability than using previous attacks.
When launching its attack the adversary introduces a
transaction of value $v_t$ in a block that is included in the public
chain and that it attempts to double spend. We also assume that the
adversary cannot break cryptographic primitives. 
Contrary to the selfish mining's adversary model~\cite{selfish14,Gervais16}, our model does not assume that the adversary has a privileged network access,
which is required in selfish mining when the adversary releases a
conflicting block it had pre-mined in reaction to the extension of the
public chain by an honest miner. For simplicity, we consider that every newly discovered and propagated block is almost instantaneously received by all miners. Several works evaluated and modeled network propagation delays in various cryptocurrencies~\cite{DeckerW13,tcao2020,cao2021characterizing}.




\section{Attack Overview}
\label{sec:overview}

This section provides a high-level description of the Dual Private Chain (DPC) attack, where an adversary maintains two private chains. It then summarizes the respective roles of adversary's two private chains and their interactions.  

\subsection{Intuition}

In a DPC attack, the adversary maintains two private chains from which it might release block headers or full blocks with the ultimate goal of double spending.
During the attack, both of the adversary's private chains compete with the public chain and may diverge from it starting from different blocks.
At a given point in time, the adversary might dedicate its full hash power to one of its private chains, or divide its hash power to simultaneously extend both private chains. 

The DPC attack starts when the adversary creates a transaction of value $v_t$ that is the basis for its double spending attempt. Once the adversary generates the block that contains this transaction, she initializes both its private chains with it and starts mining on it. Initially, the two chains are therefore equal, but they might diverge or converge again later on depending on the created blocks. 
The double spending attack succeeds if the double spending chain becomes longer than the public chain and if the public chain contains $z=6$ blocks that have been included after the block that contains the initial transaction of the adversary. 

\emph{Role of the Distraction Chain.}
The first private chain that the adversary maintains is called the \emph{distraction chain}. We present perishing mining, a strategy that the adversary employs to maintain its first private chain to waste the hash power of type 2 honest miners and slow down the public chain. Whenever the adversary divides its hash power to simultaneously mine on its two private chains, it dedicates $\beta$ of its hash power to mine on its first private chain. This chain is private in the sense that the adversary never releases the full blocks, but only the corresponding block headers. The strategy that the adversary applies on its distraction chain divides the honest miners so that they mine on different blocks, and wastes the hash power of type 2 honest miners, which collectively account for hash power $\mu$. The adversary leverages a BDOS-like attack to only share the header of blocks it discovers on the distraction chain (see Section~\ref{sec:dpc_spv}). As the body of those blocks contain adversary-created transactions that are never publicly released, only type 2 honest miners mine on them. In this way, the adversary can distract type 2 honest miners from mining on the public chain.


\emph{Role of the Double Spending Chain.}
The adversary maintains a second private chain to attempt to double spend, and we therefore call this chain the \emph{double spending chain}. Whenever the adversary is simultaneously mining on its two private chains it dedicates $\alpha(1{-}\beta)$ of the global hash power to its second private chain. This chain is private in the sense that, even though block headers might be released, the actual blocks it contains are only published if the double spending attack is successful. 
Following previous analyses~\cite{nakamoto2008bitcoin,rosenfeld2014analysis}, we consider that a double spending attempt is successful when: (i) the double spending chain's length is larger than or equal to the public chain's length; and (ii) $z\texttt{-}1$ blocks have been appended after the block that contains the adversary's initial transaction ($z=6$ in Bitcoin). 

\begin{figure}[t]
	\centering
\begin{subfigure}[b]{0.49\textwidth}
	\centering
	\includegraphics[width=\textwidth]{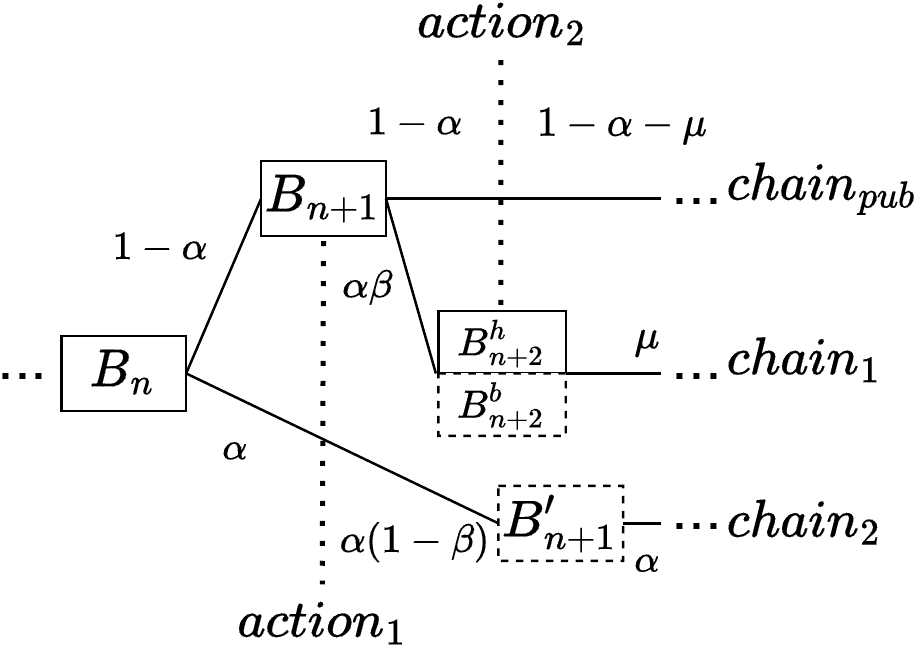}
	\caption{Case 1.}
\end{subfigure}
\hfill
\begin{subfigure}[b]{0.49\textwidth}
	\centering
	\includegraphics[width=\textwidth]{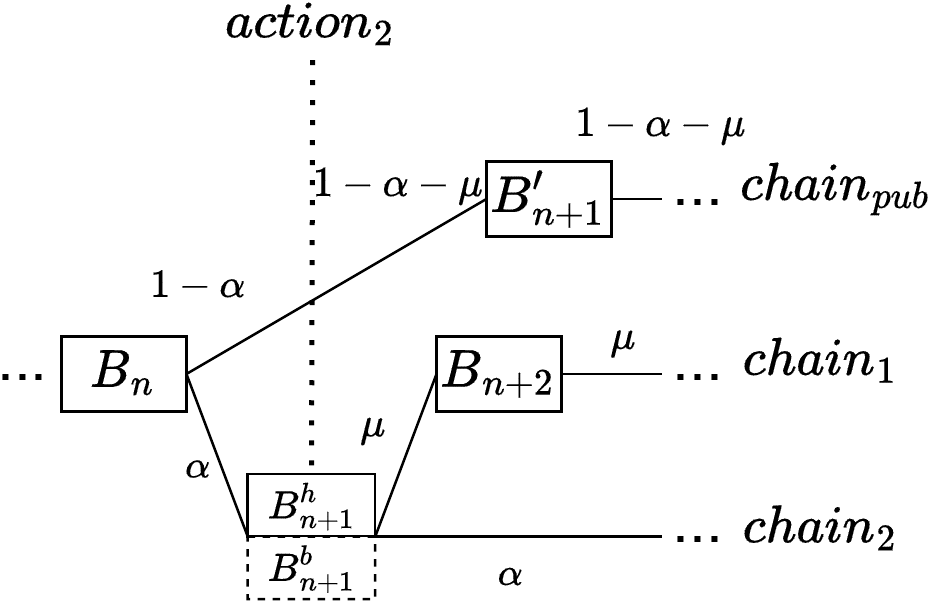}
	\caption{Case 2.}
\end{subfigure}
	\caption{Illustration of two possible cases that would lead type 2 miners to waste their hash power during a DPC attack. $B_n,B_{n+1},B_{n+1}',B_{n+2}$ are full blocks, $B_{n+1}^h,B_{n+2}^h$ are block headers, and $B_{n+1}^b,B_{n+2}^b$ are block bodies. We use solid rectangles when the content of a block is visible to honest miners, and a dotted rectangle when it is hidden by the adversary. We note interesting adversary's actions with $action_1$ and $action_2$ (see text for explanations).} 
	\label{fig:dpcm2}
\end{figure}

\subsection{Interplay Between the Two Private Chains} 

Whenever type 1 and type 2 miners are mining on the same block, the adversary divides its hash power to concurrently mine with hash power $\alpha\beta$ on the last block of its distraction chain, which is then equal to the public chain, and mine with mining power $\alpha(1-\beta)$ on its double spending chain. The adversary's goal is then to create a fork and release a block header so that type 1 and type 2 honest miners mine on different blocks. Note that the adversary will use all its hash power on the second private chain as long as the first private chain is longer than the public chain. This hash power shifting between two private chains is at the core of the DPC attack, which is detailed in Section~\ref{sec:combine_all}. 


In the DPC attack, the adversary executes different actions to lead the honest miners to mine on different blocks. Fig.~\ref{fig:dpcm2} shows two possible scenarios where the attack is initialized based on block $B_n$. The adversary generates a pair of conflicting transactions for its double spending attack. The first transaction is released to the public network and collected by the honest miners. The second transaction is kept private by the adversary. In both examples, after $action_1$, the adversary separates her hash power into two parts: she uses $\alpha\beta$ to work on public block lead $B_{n+1}$, and $\alpha(1-\beta)$ to work on extending $chain_2$ to double spend. After $action_2$ the adversary releases the block header and uses all of her hash power to extend $chain_2$ for double spending. In both cases, type 2 honest miners (with $\mu$ of global hash power) are led to generate some blocks that will never be included in the public chain due to the adversary's block body withholding strategy. Consequently, the adversary's second private chain $chain_2$, which is used to attempt to double spend, benefits from the distraction of $chain_1$. We detail the DPC attack in \S\ref{sec:dpc_spv}.

\section{The Dual Private Chains Attack}
\label{sec:dpc_spv}

This section presents the details of the DPC attack, which attempts to lure type 2 honest miners away from extending the public chain, thus, facilitates a double spending attack. We first describe perishing mining, a strategy that a miner can use to slow down the progress of the public chain by making honest miners mine on different blocks. We then describe the full DPC attack that builds on perishing mining to maintain the adversary's first private chain.
We provide an additional discussion on the DPC attack in \S\ref{sec:discussion}.

\subsection{Perishing mining}\label{pmstrategy} 


We call \textit{perishing mining} the strategy that the adversary uses on the distraction chain (whenever she is mining on it). After the initialization of the perishing mining strategy, the distraction chain and the public chain mine on the same block. The adversary's action then depends on whether the next block is discovered by the public miners or by itself, as shown in Alg.~\ref{alg:perishing} (Appx.~\ref{pm-algo}).  
First, when the adversary discovers a block $B_{n+1}$ that makes its distraction chain longer than the public chain, it releases the corresponding block header to the network (Alg.~\ref{alg:perishing}, l.~\ref{perishing2}). Upon receiving this header, type 2 miners start mining based on it, while type 1 miners continue working on block $B_n$. 
Second, when type 1 miners discover a block, the public chain is extended (Alg.~\ref{alg:perishing}, l.~\ref{perishing3}). 
Third, when type 2 miners find a block, the public chain is extended when the public chain is equal to the private chain (Alg.~\ref{alg:perishing}, l.~\ref{perishing4}). Otherwise, the block is abandoned due to the incomplete block verification, which wastes the hash power of type 2 miners. Note that when type 2 miners are optimistic the private chain is extended when it is not equal to the public chain (Alg.~\ref{alg:perishing}, l.~\ref{perishing5}).

\begin{figure}[t]
	\centering
	\begin{subfigure}[b]{0.42\textwidth}
		\centering
		\includegraphics[width=\textwidth]{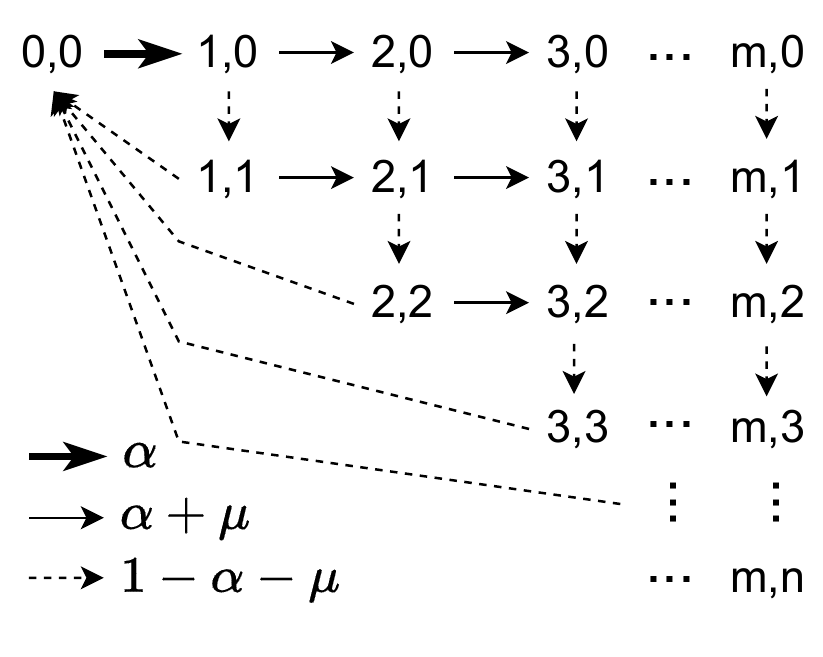}
		\caption{With optimistic type 2 miners.}
		\label{fig:mdp-spv}
	\end{subfigure}
	\hfill
	\begin{subfigure}[b]{0.42\textwidth}
		\centering
		\includegraphics[width=\textwidth]{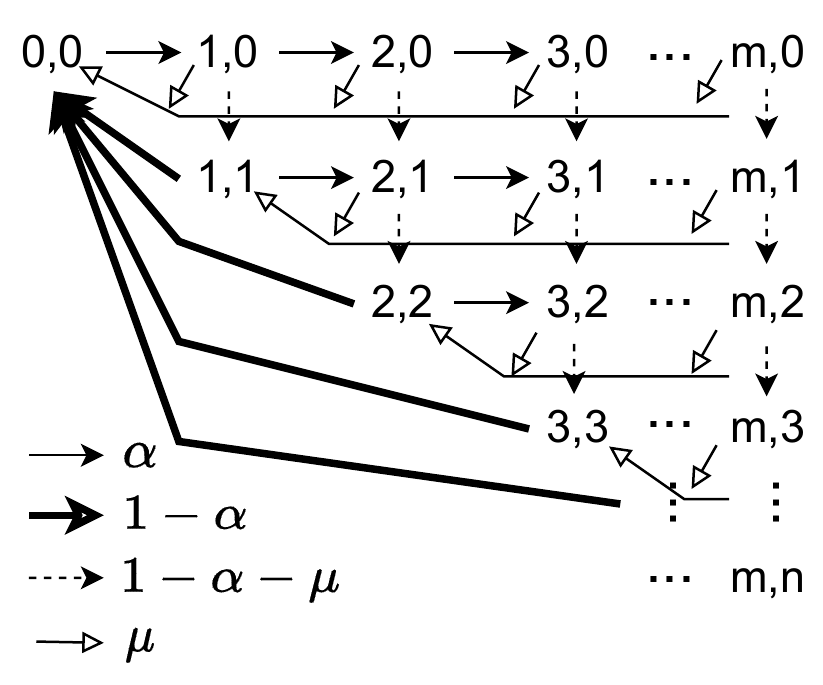}
		\caption{With pessimistic type 2 miners.}
		\label{fig:mdp-type2}
	\end{subfigure}
	
	\caption{Perishing mining's Markov chain models with optimistic and pessimistic type 2 miners. Arrows that do not lead to a state (on the right subfigure) represent the wasted mining effort of pessimistic type 2 miners. Only the top-left transition on the left figure has probability $\alpha$.}
	\label{fig:pm}
\end{figure}



Fig.~\ref{fig:pm} illustrates the MDP models of the perishing mining strategy assuming that type 2 miners are either optimistic or pessimistic. In this Markov chains, $\alpha$, $\mu$ and $1{-}\alpha{-}\mu$, are respectively the probabilities for the adversary, type 2 and type 1 miners to discover a block.
We use a tuple $(i,j)$ to denote the state in perishing mining's MDP, where $i$ and $j$ are respectively the lengths of the private chain and the public chain. The fact that the adversary adopts the public chain whenever it is longer than the private chain implies that $i {\le} j$. The adversary releases the header of the leading block to lure type 2 miners. When type 2 miners are optimistic (Fig.~\ref{fig:mdp-spv}), the adversary relies on type 2 miners that also attempt to extend the private chain. When type 2 miners are pessimistic (Fig.~\ref{fig:mdp-type2}), the adversary is not able to use them to extend the private chain. We evaluate the negative impact of perishing mining on the public chain growth in \S\ref{impact-pm}.

\subsection{Combining Perishing Mining and Double Spending}
\label{sec:combine_all}

The DPC attack leverages the perishing mining strategy to distract type 2 miners and facilitate double spending. 
Alg.~\ref{alg:bracha} (Appx.~\ref{appx:pseudocode}) details the attack's pseudocode, where $l_1$, $l_2$, and $l_{pub}$ represent the length of the first private chain $chain_1$, the second private chain $chain_2$, and the public chain $chain_{pub}$ respectively.
During the DPC attack, the two invariants $l_2 \le l_1$ and $l_{pub} \le l_1$ are verified. 
The distraction chain is therefore always the longest chain among the three chains, and can adopt the public chain and the double spending chain when it is not the longest chain. For example, if it happens that the double spending becomes the longest chain then the distraction chain is set to be equal to the double spending chain. As a consequence, the type 2 miners would mine on the headers of the double spending chain, which would facilitate the double spending attack.

When the DPC attack starts, all three chains are equal and all miners mine on the same block (Alg.~\ref{alg:bracha}, l.~\ref{line:init}). The adversary's actions are defined in reaction to block discoveries.


When the adversary finds a block on the distraction chain (Alg.~\ref{alg:bracha}, l.~\ref{line:chain1}), it releases the corresponding block header so that type 2 miners mine on it, because the distraction chain is then the longest chain. If the two private chains are equal, the newly found block also extends the double spending chain. As a consequence, the adversary extends the distraction chain, and 
type 1 miners mine on the last full block of the public chain while type 2 miners mine on the last block header of the distraction chain. The adversary then allocates all its hash power ($\alpha$) to mining on the double spending chain. 	

When the adversary finds a block on its double spending chain (Alg.~\ref{alg:bracha}, l.~\ref{line:chain2}), it releases the block header if the second private chain becomes the longest chain. In this case, type 2 miners then mine on the double spending chain. The first private chain also adopts the second private chain so that the total hash power used to extend the double spending chain is $\alpha+\mu$. When the second private chain is shorter than the public chain, the adversary keeps mining on it with $1-\beta$ of its hash power. As soon as the double spending chain becomes longer than the public chain and that at least 6 blocks have been appended to the public chain since the beginning of the attack, the adversary uses the double spending chain to override the public chain, and the DPC attack succeeds.

When type 1 miners find a block (Alg.~\ref{alg:bracha}, l.~\ref{line:honest}), they extend the public chain. If the public chain becomes the longest chain, then all honest miners will mine on the public chain and the adversary modifies its first private chain so that it adopts the public chain. The adversary then allocates $\alpha\beta$ of hash power to its distraction chain so that it tries to generate a block that will divide again the honest miners.  

When type 2 miners find a block (Alg.~\ref{alg:bracha}, l.~\ref{line:spv}), three cases are possible. First, the double spending chain is extended if two private chains are equal and longer than the public chain. Second, the public chain and first private chain are extended if they are equal. Finally, in the other cases the newly discovered block is abandoned, which wastes the hash power of type 2 honest miners. The DPC attack can be tailored to optimistic or pessimistic type 2 miners. For example,  lines~\ref{line:spv1},~\ref{line:spv2},~\ref{line:spv3},~\ref{line:spv4},~\ref{line:spv5} in Alg.~\ref{alg:bracha} describe the adversary's behavior with optimistic type 2 honest miners.  

\subsection{Markov Decision Process of the DPC Attack} 

We establish the Markov decision process (MDP) of the DPC attack by simultaneously considering the two private chains and observing that each state is a 5-tuple $(l_{pub}, l_1, l_2, s_{(pub,1)}, s_{(1,2)})$. $l_{pub}$, $l_1$, and $l_2$ are respectively the lengths of the public chain $chain_{pub}$, the first private chain $chain_1$, and the second private chain $chain_2$. $s_{(pub,1)}, s_{(1,2)} \in \{\texttt{t(rue)}, \texttt{f(alse)}\}$ respectively indicate whether $chain_{pub}$ is equal to $chain_1$, and whether $chain_1$ is equal to $chain_2$. 


Based on the relations between the three chains (synchronized or not), we identified 10 types of states in the presence of optimistic type 2 miners, and 9 types of states in presence of pessimistic type 2 miners. The corresponding transitions are presented in Tables~\ref{transition} and~\ref{transition-type2} (in Appendix~\ref{appx:pseudocode}). Note that we were not able to obtain closed form formulas for the probabilities of each possible state due to the complexity of the DPC attack's MDP model. Nevertheless, we use Monte Carlo simulations to estimate the adversary's success rate and revenue, as in previous block withholding attacks~\cite{selfish14,Gervais16,stubborn16}.



Case 0 is the initial state of the attack. Case 4 captures the attack success, which happens if the public and the double spending chains contain more than 6 blocks, and if the double spending chain is longer than the public chain. Cases 1.x, 2.x, 3.x are all possible intermediary states and consider scenarios that differ based on the lengths of the chains, and whether or not they are equal, which happens when the adversary reinitializes one or both of its private chains. 

We emphasize that an adversary that executes the DPC attack earns a mining reward only when the double spending chain succeeds. In this case, the adversary earns the block mining reward that corresponds to the private blocks it mined that end up in the public chain and the value of the transaction it managed to double spend. We use $v_b$ for the value of blocks, and $v_t$ for the value of the double spent transactions.

\section{Analysis using Monte Carlo Simulations}
\label{sec:perf_eval}

This section evaluates the perishing mining strategy and the DPC
attack using Monte Carlo simulations that react based on the event of
block discovery.

\subsection{Methodology and Settings}

We evaluate perishing mining and the DPC attack using random walks in their respective Markov decision processes. Our evaluations are based on Python scripts. In each scenario, we simulate the creation of 2,016 blocks, repeat each configuration 10,000 times, and report the average of the metrics of interest. 
Simulating the creation of 2,016 blocks maintains the mining difficulty constant during the experiment since Bitcoin's mining difficulty is adjusted every 2,016 blocks. 
We quantify the impact of perishing mining on the public chain's growth rate, and then evaluate the double spending success rate of the DPC attack. We compare the success rate of the DPC attack to the success rate of the classical double spending attack using the success rate formulas that were obtained by Nakamoto~\cite{nakamoto2008bitcoin} and Rosenfeld~\cite{rosenfeld11}. 
We study the various strategies with $\alpha, \mu \in [0,0.1,0.2,0.3,0.4,0.5]$ and $\beta \in [0,1]$ (by 0.01 steps). Moreover, we analyze the adversary's expected revenue in Appx.~\ref{profit}.   

%

\subsection{Impact of Perishing Mining on Chain Growth}\label{impact-pm}


\begin{figure}[t]
	\centering
	\begin{subfigure}[b]{0.48\textwidth}
		\centering
		\includegraphics[width=\textwidth]{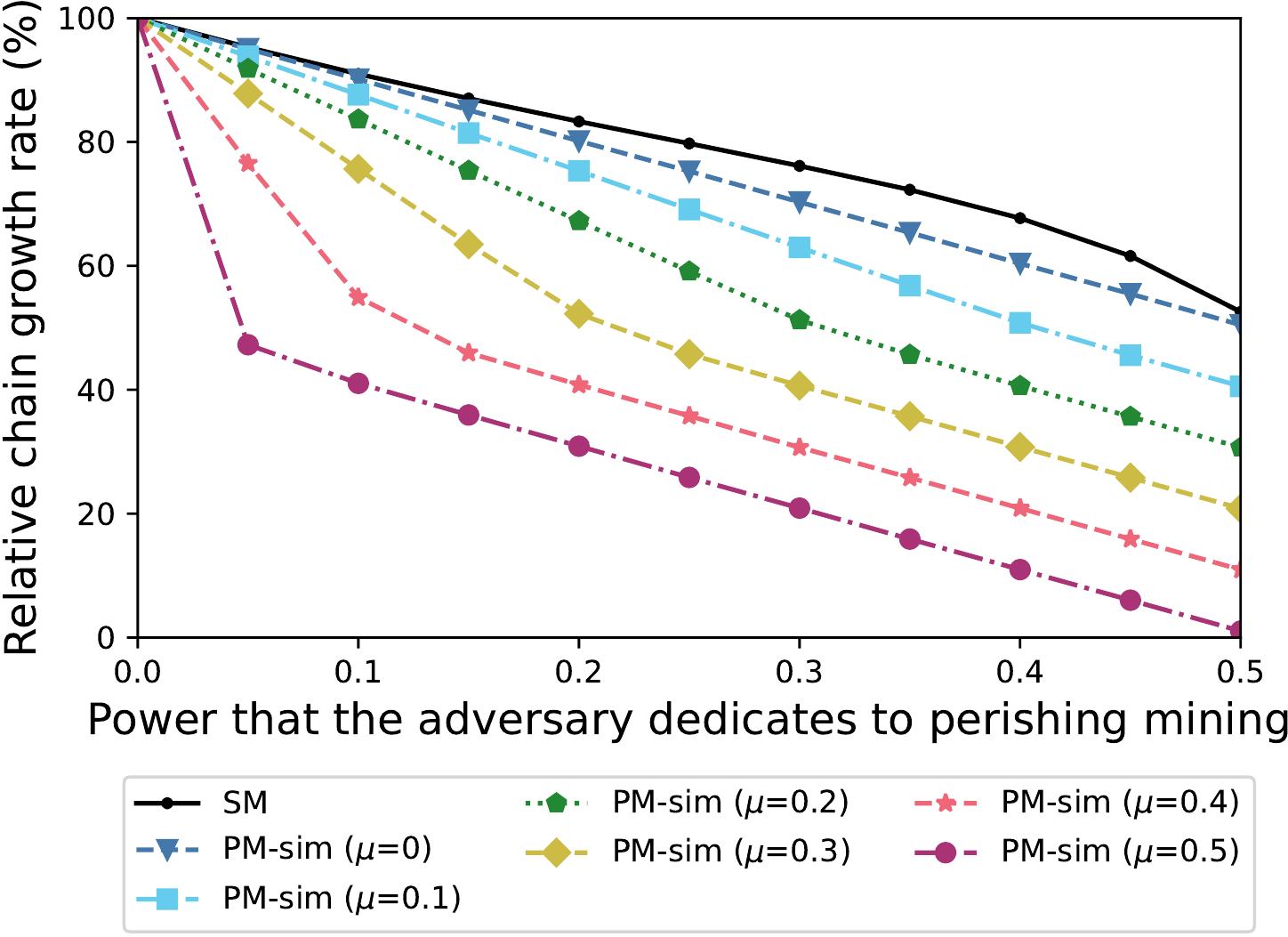}
		\caption{With optimistic type 2 miners.}
		\label{fig:pm-cgr-1}
	\end{subfigure}
		\hfill
    \begin{subfigure}[b]{0.48\textwidth}
		\centering
		\includegraphics[width=\textwidth]{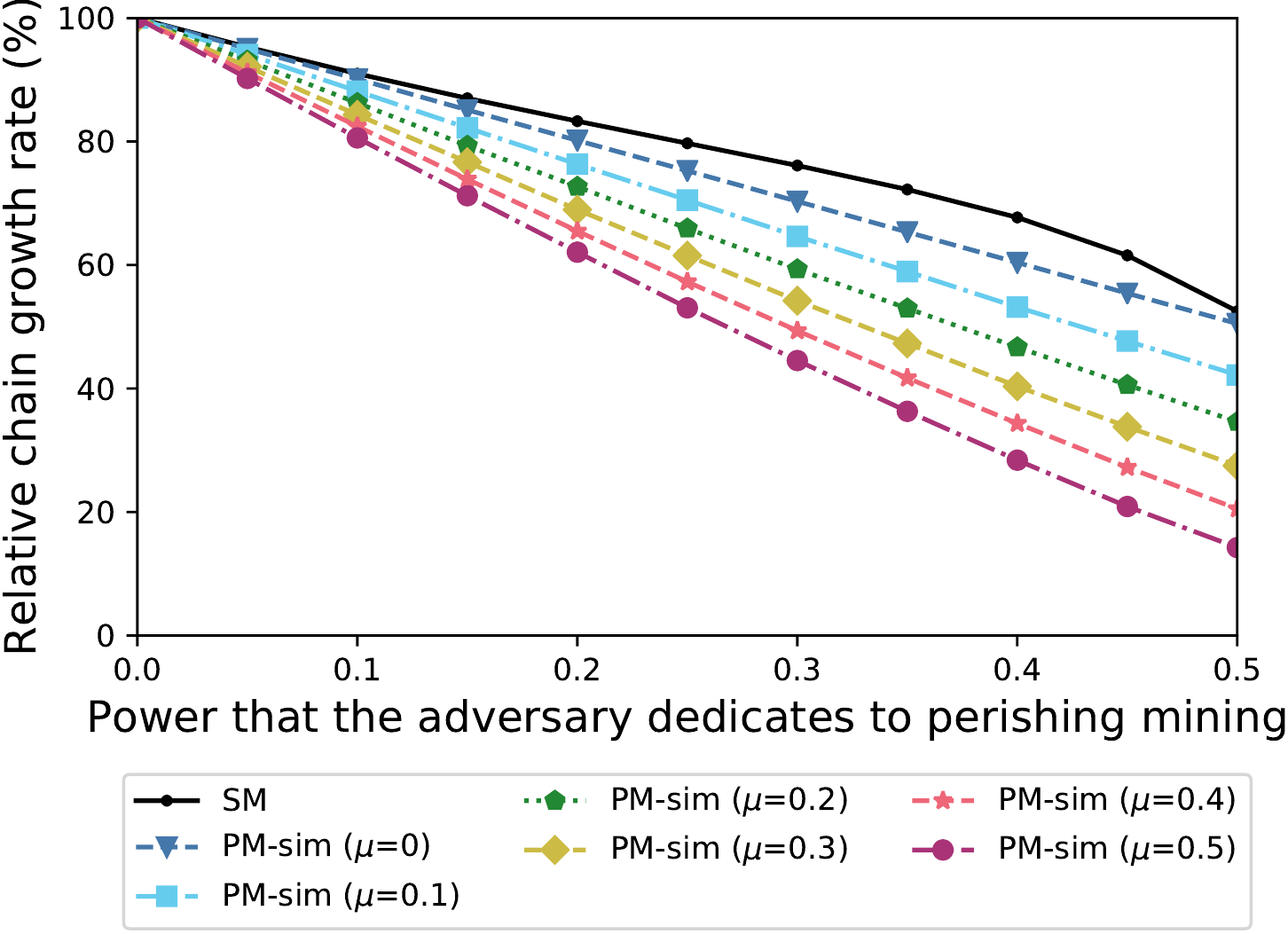}
		\caption{With pessimistic type 2 miners.}
		\label{fig:pm-cgr-2}
	\end{subfigure}
	\caption{Relative growth rate of the public chain (compared to the attack-free case) when the adversary uses selfish mining (SM) or perishing mining (PM), where type 2 miners own a fraction $\mu$ of the global power.}
	\label{fig:pmchaingrowth}
\end{figure}

In a DPC attack, the adversary leverages perishing mining strategy to inhibit public chain's growth. We now consider a scenario where the adversary constantly dedicates a fraction of its full hash power to perishing mining, so that we can quantify its effect on the growth rate of the public chain.  

Fig.~\ref{fig:pmchaingrowth} represents the relative public chain growth rate of a system under attack, which is expressed as a fraction (in \%) of the public chain growth rate in the attack-free case. We compare perishing mining to selfish mining and vary the global hash power $\mu$ of type 2 miners $0$ to $0.5$ (i.e., ranging from 0\% to 50\% of the global hash power). The public chain is extended at a lower rate when the adversary's power increases and when the global power of type 2 miners increases. By comparing Fig.~\ref{fig:pm-cgr-1} and Fig.~\ref{fig:pm-cgr-2}, one can see that perishing mining has a stronger impact with optimistic type 2 miners than with pessimistic type 2 miners, as one could expect. 

\subsection{Double Spending Success Rate}\label{sr-spv}



\begin{figure}[t]
	\centering
	\begin{subfigure}[b]{0.48\textwidth}
		\centering
		\includegraphics[width=\textwidth]{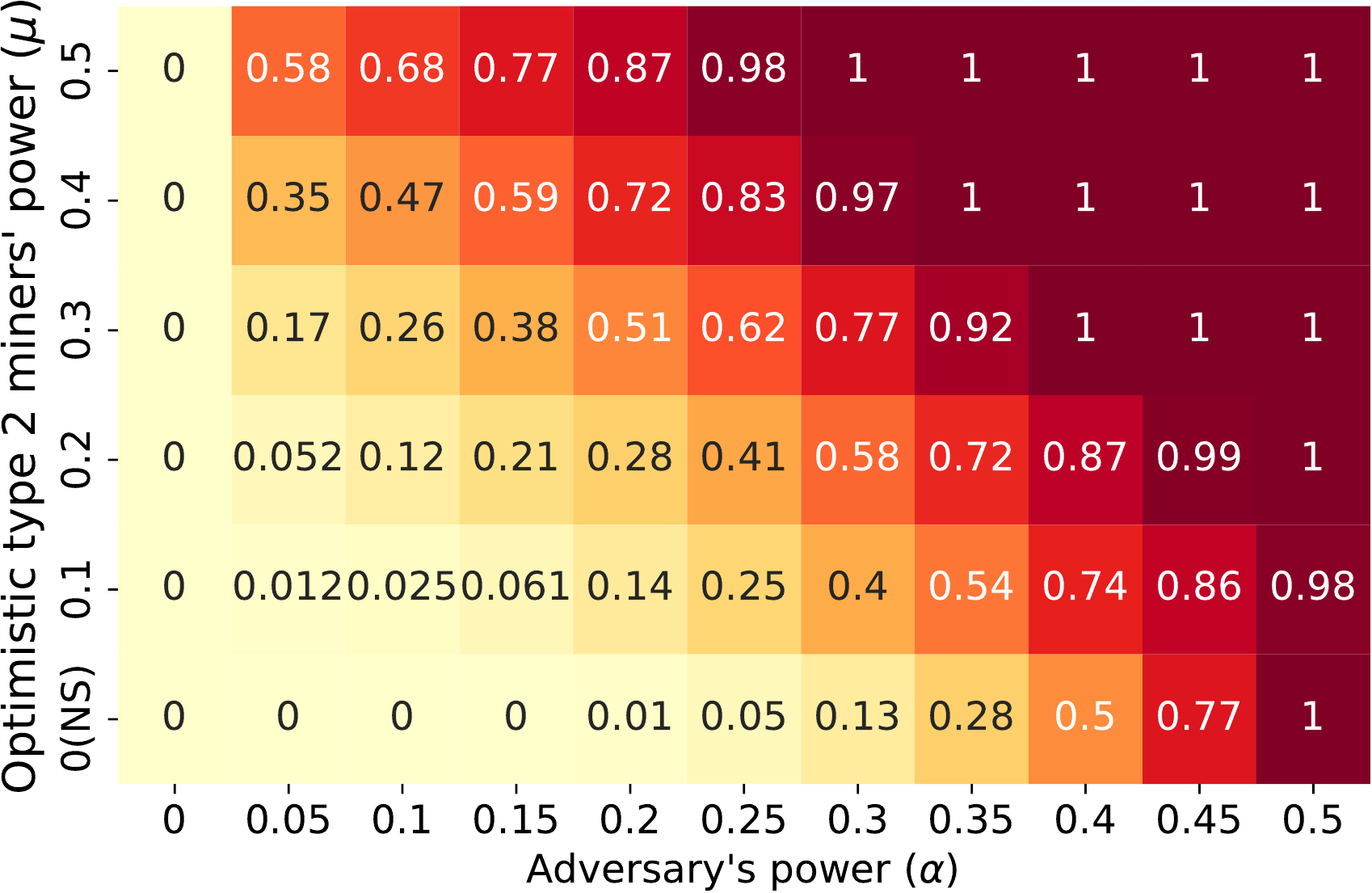}
		\caption{With optimistic type 2 miners.}
		\label{fig:successrate1}
	\end{subfigure}
	\hfill
	\begin{subfigure}[b]{0.48\textwidth}
		\centering
		\includegraphics[width=\textwidth]{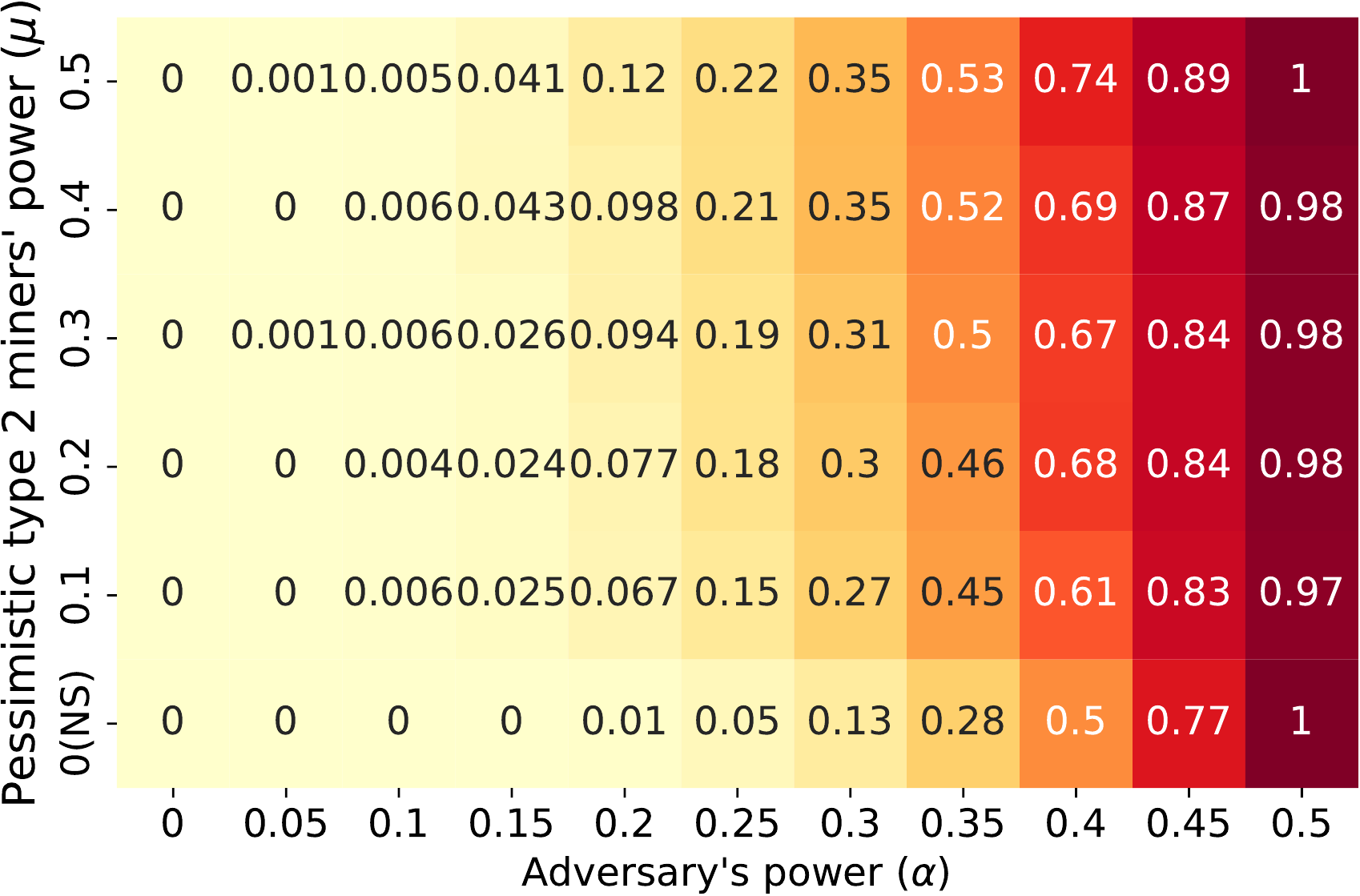}
		\caption{With pessimistic type 2 miners.}
		\label{fig:successrate2}
	\end{subfigure}

	\caption{Success rate of the DPC attack depending on the hash power $\mu$ of the type 2 miners with the best value of parameter $\beta$ within 2016 blocks. The ``NS'' line represents the success rate of the classical double spending attack (based on Nakamoto's evaluation). A darker color indicates a higher success probability.}
	\label{fig:successrate}

\end{figure}

Fig.~\ref{fig:successrate} illustrates the success rates of the DPC attack for different $\mu$ and with the best $\beta$ value that we obtained experimentally. It is interesting to observe the differences between the partitions corresponding to a given $\mu$ with the best $\beta$ value to see that maintaining distraction chain and double spending chain simultaneously makes a real difference. An adversary would be able to determine the best $\beta$ after estimating $\mu$, as we discuss in \S\ref{sec:discussion}.

In presence of type 2 miners (i.e., $\mu>0$), the DPC attack's success rate is always higher than the one of the traditional double spending attack (i.e., 0(NS) in Fig.~\ref{fig:successrate}). The success rate of the double spending attack (with 6 confirmations) with $\alpha=0.2$ (the power of the biggest mining pool) increases from 1\% to 87\% (or from 1\% to 12\%) via the DPC attack depending on $\mu$ as shown in Fig.~\ref{fig:successrate1} (or Fig.~\ref{fig:successrate2}). The impact of optimistic type 2 miners on DPC attack' success rate is more severe than pessimistic type 2 miners, for example, if $\mu=0.2$ and $\alpha=0.2$, the DPC attack's success rate is $28\%$ in Fig.~\ref{fig:successrate1} while it is $7.7\%$ in Fig.~\ref{fig:successrate2}. 

Importantly, the DPC attack lowers Bitcoin's safety bound, i.e., the minimum hash power that the adversary needs to double spend or break the chain's consistency. 
For instance, when $\mu=0.5,0.4,0.3,0.2$ and type 2 miners are optimistic, a DPC adversary with $30\%,35\%,40\%,45\%$ of the global hash power could completely manipulate the blockchain (i.e., 100\% success rate in Fig.~\ref{fig:successrate1}), which is more threatening than the existing block withholding attacks~\cite{selfish14,Gervais16,stubborn16}. 

\begin{figure}[t]
	\centering
	\begin{subfigure}[b]{0.48\textwidth}
		\centering
		\includegraphics[width=\textwidth]{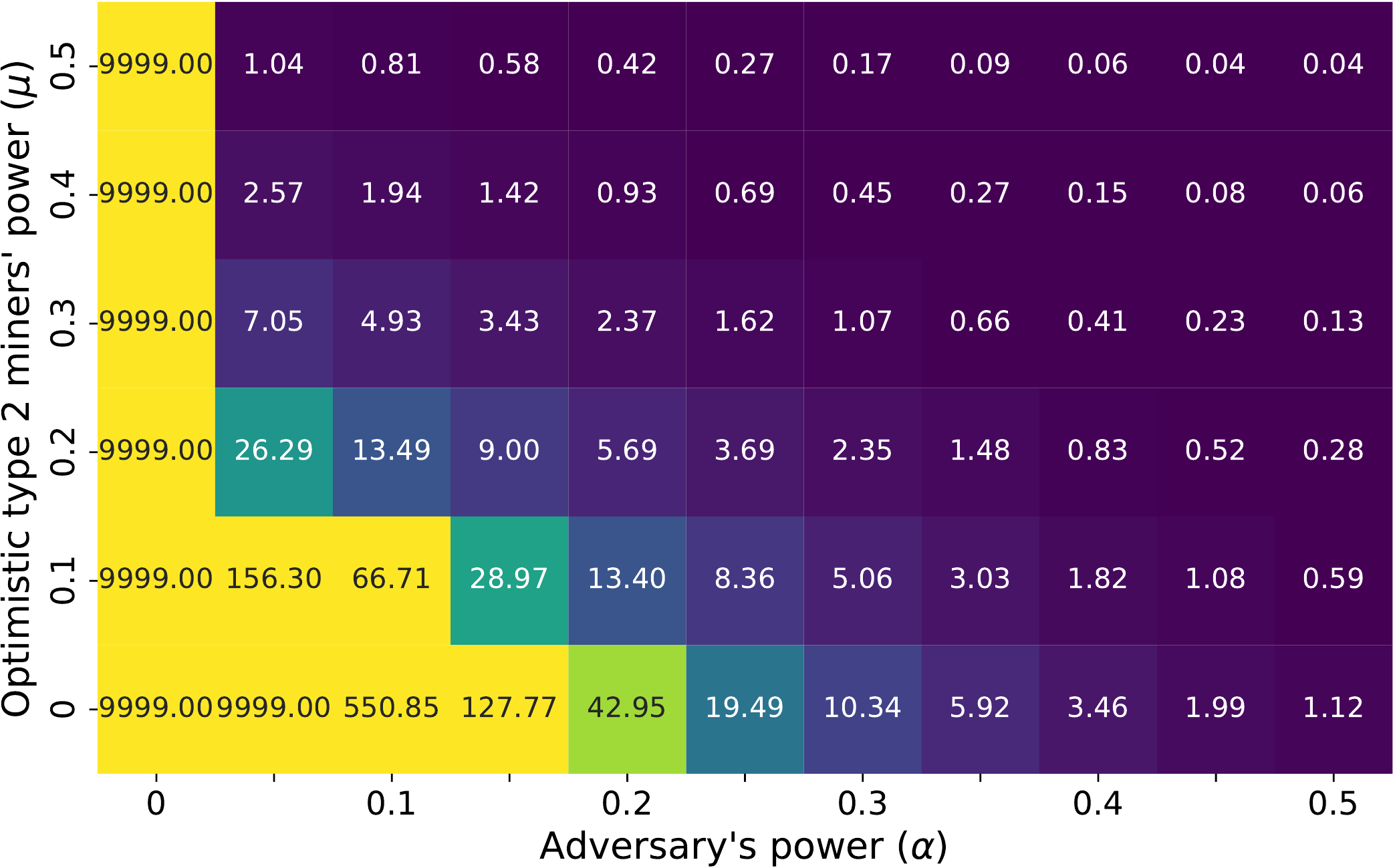}
		\caption{With optimistic type 2 miners.}
		\label{fig:vt1}
	\end{subfigure}
	\hfill
	\begin{subfigure}[b]{0.48\textwidth}
		\centering
		\includegraphics[width=\textwidth]{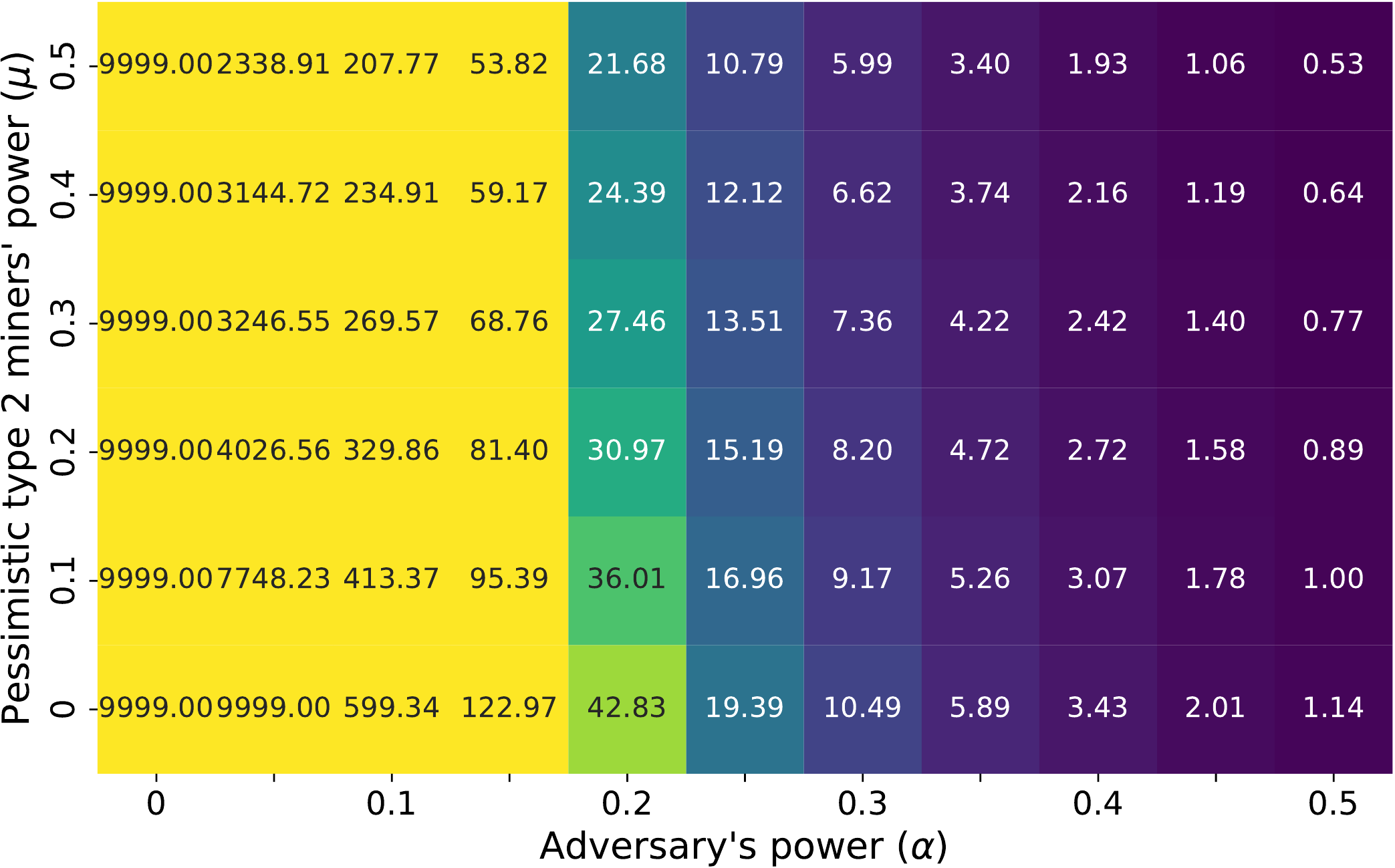}
		\caption{With pessimistic type 2 miners.}
		\label{fig:vt2}
	\end{subfigure}
	
	\caption{Minimum value for $\frac{v_t}{v_b}$ for the DPC attack to be more profitable than honest mining depending on $\mu$. ``9999'' represents $\frac{v_t}{v_b} \geq 9999$. }
	\label{fig:vt}
\end{figure}

Inspired by M. Rosenfeld~\cite{rosenfeld11}, we further evaluate the safe transaction value (i.e., the suggested maximum value of transaction for clients) against double spending attack. Fig.~\ref{fig:vt} plots the minimum value for $\frac{v_t}{v_b}$ that allows the DPC attacker to be more profitable than honest mining. When $\mu=0.2$ and type 2 miners are optimistic, the adversary with $0.05,0.1,0.15,0.2$ (the possible hash power share of mining pools in Bitcoin) of global hash power is incentivized to perform DPC attack as long as the merchants are willing to accept the transaction with $26.29*v_b,13.49*v_b,9*v_b,5.69*v_b$ BTC respectively (as shown in Fig.~\ref{fig:vt1}). In the same case, when type 2 miners are pessimistic, the safe transaction value would increase and become $4026.56*v_b,329.86*v_b,81.4*v_b,30.97*v_b$.    
Bitcoin's future block reward halving will decrease both the threshold to launch profitable DPC attacks and the safe transaction value, which confirms Carlsten et al.'s previous observation~\cite{carlsten2016instability}.


\section{Attack Discussion}
\label{sec:discussion}


\textbf{Attack variants.}
We have presented the DPC attack we found to be the most effective when 
the adversary splits its hash power in two constant parts $\alpha\beta$ and $\alpha(1-\beta)$.
We foresee that one could devise variants of the DPC attack, e.g., using techniques that have been applied to selfish mining~\cite{karame2012doublespending,selfish14,rosenfeld2014analysis,sapirshtein2015optimal}. 
In these variants the adversary would mine on different blocks depending on the system's state, or dedicate a different fraction of its hash power to extend each of its two private chains. We leave the study of these variants to future work.  


\textbf{Estimating $\mu$ and Selecting $\beta$.}
It is sufficient for the adversary to approximate the value of $\mu$, which is the proportion of the global hash power that belongs to type 2 miners, for a DPC attack to be successful, as our experimental results demonstrate. However, in practice, an adversary would be able to optimize its DPC attack by determining a precise value for $\mu$. The adversary can estimate $\mu$ based on some public websites~\cite{f2pool}, or establish direct connections with the public mining pools to perform a statistical analysis. Moreover, the perishing mining strategy that we present in this paper can be used as a probing technique to measure $\mu$. Indeed, the adversary can directly monitor the impact of perishing mining on the public chain and compute $\mu$ based on its growth rate. 
Once the exact value of $\mu$ is known, an adversary can find the best $\beta$ for the DPC attack by replicating our experiments.

\textbf{Attack Detection and Prevention.}
The DPC attack leverages the fact that type 2 miners, which include SPV miners, accept block headers without waiting for and verifying the corresponding transactions. 
One partial countermeasure against the DPC attack would consist in miners deliberately choosing to stop mining on block headers alone.
However, it does not seem reasonable to assume that all miners would avoid this strategy because they can start working on the next block earlier than other miners and therefore increase their profit.  
Type 2 miners could also avoid mining on the adversary's blocks by accepting to mine only on blocks that were discovered from known mining pools. It is unclear whether this modification would have undesired security implications, e.g., regarding the decentralization of proof-of-work blockchains, or because pool sub-miners run a mining software that is developed internally and independently from the official protocol specification~\cite{eyal2015miner}. 
In addition, this modification would require type 2 miners to trust mining pools, and a malicious pool manager would still be able to execute the DPC attack. 

Another idea would be for type 2 miners to 
stop mining on a block header if the associated transactions are not received before a maximum delay and then mine on the last full block. However, the adversary could also update its strategy to regularly send the unmatched block bodies so that type 2 miners keep mining on its blocks. It is unclear whether this countermeasure would be efficient, and in particular in practical settings. 
Moreover, the variation of message delays in Bitcoin's peer to peer network would sometimes lead type 2 miners to reject blocks that are generated by honest miners, and might imply possible DoS attacks. 

\textbf{Evidence of type 2 mining.} In practice, it is difficult to know the exact strategy that miners follow. However, previous works have provided evidence of SPV mining~\cite{mirkin2020bdos,emptyblocks,halfspv,f2pool,spvpools}. 
Our assumptions in this work are not stronger since our pessimistic type 2 miners are more conservative than SPV miners. 
In 2020, 9+ mining pools representing 36\% of the global power produced empty blocks, which one might consider evidence of SPV mining~\cite{theblock}. We analysed the Bitcoin blockchain and found that Antpool, Binance, F2pool, Huobi, Poolin, ViaBTC published empty blocks from 01/2021 to 02/2022 and collectively represent more than 60\% of the global power.



\section{Conclusion}
\label{sec:conclusion} 

In this paper, we proposed perishing mining, a novel adversarial mining strategy that slows down the public chain by leveraging the verifier's dilemma. 
We then described the dual private chain (DPC) attack where an adversary dedicates a part of its hash power to the perishing mining strategy and launches a parallel double spending attack.
We established the Markov decision process of both the perishing mining and the DPC attack. We relied on Monte Carlo simulations to quantify the impact of perishing mining on the public chain growth, and evaluate the double spending success rate of the DPC attack. 
Our performance evaluation showed that the DPC attack is more powerful than the classical double spending attack as soon as a fraction of the miners mine on blocks without verifying their transactions. 
We also evaluated the revenue an adversary could expect from running the DPC attack, and showed that an adversary with sufficient funds or with sufficient hash power would maximize its revenue with the DPC attack.

\bibliographystyle{splncs04}  
\bibliography{biblio}

\appendix

\begin{subappendices}
\renewcommand{\thesection}{\Alph{section}}%

\newpage
\section{Pseudocode of Perishing Mining}\label{pm-algo}

Alg.~\ref{alg:perishing} details the pseudocode of perishing mining with either optimistic or pessimistic type 2 miners. Perishing mining aims at slowing down the growth of the public chain.  

\begin{algorithm}[hbpt]
	\caption{\textit{The perishing mining strategy}.} 
	\label{alg:perishing}
	\begin{algorithmic}[1]
		\scriptsize 
		
		\State \textbf{function} \texttt{Init()} 
		\State \hspace*{2em} $chain_{pub}$, $chain_{1}$ $\leftarrow$ publicly known blocks \comalgo{public and private chains}
		\State \hspace*{2em} $l_{pub} = l_1 = 0$
		\State \hspace*{2em} \textbf{mine} on $chain_{1}$'s head
		\State
		
		\State \textbf{upon event} \texttt{Adv. finds block B on $chain_{1}$} \textbf{do} \comalgo{prob. $\alpha$}
		\State \hspace*{2em} Adv. appends B to $chain_1$
		\State \hspace*{2em} $l_1\texttt{++}$
		\State \hspace*{2em} \textbf{release} Header(B) \comalgo{withhold block body} \label{perishing2}
		\State \hspace*{2em} \textbf{mine} on B
		\State
		
		\State \textbf{upon event} \texttt{Type 1 miners find and append block B on $chain_{pub}$} \textbf{do} \comalgo{prob. $1-\alpha-\mu$}
		\State \hspace*{2em} $l_{pub}\texttt{++}$ \label{perishing3}
		\State \hspace*{2em} \textbf{if} $l_{pub} > l_1$ \textbf{then}
		\State \hspace*{4em} \texttt{Init()} \comalgo{adopt} \label{perishing1}
		\State
		
		\State \textbf{upon event} \texttt{Type 2 miners find block B} 
		\textbf{do} \comalgo{prob. $\mu$}
		\State \hspace*{2em} \textbf{if} $chain_{pub} \texttt{==} chain_1$ \textbf{then} 
		\State \hspace*{4em} Type 2 miners append B to $chain_{pub}$		
		\State \hspace*{4em} $\texttt{Init()}$ \comalgo{adopt} \label{perishing4}
		\State \hspace*{2em} \textbf{else}  
		\State \hspace*{4em} $\{ \texttt{Init()} \}$ \comalgo{pessimistic type 2 miners.}
		\State \hspace*{6em}  or 
		\State \hspace*{4em} $\{$Type 2 miners append B to $chain_1$ \comalgo{optimistic type 2 miners}
		\State \hspace*{4em} $l_1\texttt{++}$  \label{perishing5}
		\State \hspace*{4em} \textbf{mine} on B$\}$
	\end{algorithmic}
\end{algorithm}




\section{Pseudocode, States and Transitions of the DPC Attack}
\label{appx:pseudocode}

Alg.~\ref{alg:bracha} presents the pseudocode of the full DPC attack where an adversary manages two private chains to attempt to double spend. 
Table~\ref{transition} details the states that can be reached during a DPC attack and the possible transitions between them in presence of optimistic type 2 miners. Table~\ref{transition-type2} assumes the presence of pessimistic type 3 miners. Transitions happen when a block is discovered either by the adversary or by an honest miner (type 1 or type 2).  

\begin{algorithm}[!htbp]
	\caption{ \textit{The DPC attack, which the adversary can configure to assume that all type 2 miners are pessimistic or optimistic.}} 
	\label{alg:bracha}
	\begin{algorithmic}[1]
		\scriptsize 
		
		\State \textbf{function} \texttt{Init} \textbf{do} \label{line:init}
		\State \hspace*{2em} $chain_{pub}$ $\leftarrow$ publicly known blocks 
		\State \hspace*{2em} $chain_{1}$ $\leftarrow$ publicly known blocks
		\comalgo{1st priv. chain (distraction chain)} 
		\State \hspace*{2em} $chain_{2}$ $\leftarrow$ publicly known blocks
		\comalgo{2nd priv. chain (double spending chain)}
		\State \hspace*{2em} \textbf{mine} on $chain_{1}$'s head with power $\alpha\beta$
		\State \hspace*{2em} \textbf{mine} on $chain_{2}$'s head with power $\alpha(1-\beta)$
		\State \hspace*{2em} $l_{pub} = l_1 = l_2 = 0$
		
		\State \textbf{upon event} \texttt{Adv. finds a block B on $chain_{1}$} \textbf{do} \label{line:chain1}
		\State\hspace*{2em} \textbf{if} $chain_{1} {\neq} chain_2$ \textbf{then}
		\State\hspace*{4em} Adv. appends B to $chain_1$ 
		\State\hspace*{4em} $l_1\texttt{++}$ 
		\State\hspace*{4em} \textbf{release} Header(B) 
		\State\hspace*{4em} \textbf{mine} on $chain_2$ with power $\alpha$ 
		\State\hspace*{2em} \textbf{else if} $chain_{1} \texttt{==} chain_2
		$ \textbf{then} 
		\State\hspace*{4em} Adv. appends B to $chain_1$ (and $chain_2$) \label{linea}
		\State\hspace*{4em} $l_1\texttt{++}$ 
		\State\hspace*{4em} $l_2\texttt{++}$ 
		\State\hspace*{4em} \textbf{release} Header(B) 
		\State\hspace*{4em} \textbf{mine} on $chain_2$ with power $\alpha$ \label{lineb}
		
		
		\State \textbf{upon event} \texttt{Adv. finds a block B on $chain_{2}$} \textbf{do} \label{line:chain2}
		\State\hspace*{2em} \textbf{if} $chain_{1} {\neq} chain_2$ \textbf{then}
		\State\hspace*{4em} Adv. appends B to $chain_2$ 
		\State\hspace*{4em} $l_2\texttt{++}$ 
		\State\hspace*{4em} \textbf{release} Header(B) 
		\State\hspace*{4em} \textbf{mine} on $chain_2$ with $\alpha(1-\beta)$  
		\State\hspace*{2em} \textbf{else if} $chain_{1} \texttt{==} chain_2 $ \textbf{then}
		\State\hspace*{4em} Execute lines~\ref{linea} to~\ref{lineb} 
		\State\hspace*{2em} \textbf{if} $l_2 \geq l_{pub} \geq 6$ \textbf{then} 
		\State \hspace*{4em} \textbf{override} $chain_{pub}$ with $chain_2$ \comalgo{double spending}

		\State \textbf{upon event} \texttt{type 1 honest miners find and append block B on $chain_{pub}$} \textbf{do} \label{line:honest}
		\State \hspace*{2em} \textbf{if} $l_{1} \texttt{==} l_{pub}$ \textbf{then}
                \State \hspace*{4em} $chain_{1} = chain_{pub}$
                \comalgo{reinit. 1st priv. chain} \label{linec} \State
                \hspace*{4em} $l_{pub}\texttt{++}$ \State
                \hspace*{4em} $l_1\texttt{++}$ \State \hspace*{4em}
                \textbf{mine} on $chain_1$ with power
                $\alpha\beta$ \label{lined}
		
		\State \hspace*{2em} \textbf{else if} $l_{1} \texttt{>} l_{pub}$ \textbf{then}
		\State \hspace*{4em} $l_{pub}\texttt{++}$
		
		
		\State \textbf{upon event} \texttt{type 2 honest miners find block B} \textbf{do}\label{line:spv}
		
		\State \hspace*{2em} \textbf{if} $chain_{1} \texttt{==} chain_{2} {\neq} chain_{pub}$ \textbf{then} 
		\State \hspace*{4em} $\{$type 2 honest miners append B to $chain_{2}$ \comalgo{conservative approach with pessimistic type 2 miners}
		\State \hspace*{4em} $chain_1 \texttt{=} chain_{pub}$
		\State \hspace*{4em} $l_2\texttt{++} \}$
		\State \hspace*{6em} or \comalgo{optimistic type 2 miners}
		\State \hspace*{4em} $\{$type 2 honest miners append B to $chain_{2}$ (and $chain_1$)~\label{line:spv1}
		\State \hspace*{4em} Execute lines~\ref{linea} to~\ref{lineb}$\}$~\label{line:spv2}

		\State \hspace*{2em} \textbf{else if} $chain_{1} \texttt{==} chain_{pub}$ \textbf{then} 
		\State \hspace*{4em} type 2 honest miners append B to $chain_1$ (and $chain_{pub}$)
		\State \hspace*{4em} Execute lines~\ref{linec} to~\ref{lined} 
		
		\State \hspace*{2em} \textbf{else} 
		\State \hspace*{4em} $\{chain_1 \texttt{=} chain_{pub}$\comalgo{conservative approach with pessimistic type 2 miners}
		\State \hspace*{4em} $l_1\texttt{=}l_{pub}\}$
		\State \hspace*{6em} or \comalgo{optimistic type 2 miners}
		\State \hspace*{4em} $\{$type 2 honest miners append B to $chain_1$~\label{line:spv3}
		\State \hspace*{4em} $l_1\texttt{++}$~\label{line:spv4} 
		\State \hspace*{4em} \textbf{mine} on $chain_2$ with power $\alpha\}$~\label{line:spv5}
		
		%
		%
		%
	\end{algorithmic}
\end{algorithm}

\begin{table}[htbp]
	\centering
	\scriptsize
	\caption{States and transitions of the DPC attack with optimistic type 2 miners (e.g., SPV miners). We use $(r_a,r_h)$ to denote the revenue of the adversary and other miners, $v_b$ to denote the value of a single block, and $v_t$ to denote the value of the double-spent transactions. Because type 2 miners could help to extend the double spending chain, we use $n_{t2}$ to denote the number of blocks that were discovered by type 2 miners on the 2nd private chain.} 
	
	\renewcommand{\arraystretch}{1.2}
	\begin{adjustbox}{angle=270}
	\begin{tabular}{p{4.6cm}|p{4.4cm}|p{1cm}|l|p{2.8cm}}
		
		\textbf{State S} & \textbf{Event} & \textbf{Prob.} & \textbf{Destination state} & \textbf{Revenue} \\ 
		$(l_{pub}, l_1, l_2, {sync}_{(pub,1)}, {sync}_{(1,2)})$ & & & & $(r_a,r_h)$ \\
		\hline
		
		Case 0 (Contains Init state) &  Adv. extends $chain_1 {=} chain_2$ & $\alpha$ & $(l_{pub},l_1+1,l_2+1,\texttt{f},\texttt{t})$ &  $(r_a,r_h)$\\ \cline{2-5}
		$(l_{pub}, l_1, l_2, \texttt{t}, \texttt{t})$ & Type 1 or 2 extends $chain_{pub}$ & $1-\alpha$ & $(l_{pub}+1,l_1+1,l_2,\texttt{t},\texttt{f})$) & $(r_a,r_h+v_b)$ \\ 
		\hline
		
		Case 1.1 & Adv. or type 2 extends $chain_1 {=} chain_2$ & $\alpha+\mu$ & $(l_{pub}, l_1+1, l_2+1, \texttt{f}, \texttt{t})$ & $(r_a,r_h)$ \\ \cline{2-5}
		$(l_{pub}, l_1 > l_{pub}, l_2, \texttt{f}, \texttt{t})$ & Type 1 extends $chain_{pub}$ & $1-\alpha-\mu$ & $(l_{pub}+1, l_1, l_2, \texttt{f}, \texttt{t})$ & $(r_a,r_h+v_b)$\\ 
		\hline
		
		Case 1.2 & Adv. or type 2 extends $chain_{1}$ & $\alpha+\mu$ & $(l_{pub}, l_1+1, l_2+1, \texttt{f}, \texttt{t})$ & $(r_a,r_h)$ \\ \cline{2-5}
		$(l_{pub}, l_1=l_{pub}, l_2, \texttt{f}, \texttt{t})$ & Type 1 extends $chain_{pub}$ & $1-\alpha-\mu$ & $(l_{pub}+1, l_1+1, l_2, \texttt{t}, \texttt{f})$ & $(r_a,r_h+v_b)$  \\ 
		\hline
		
		Case 2.1 & Adv. extends $chain_1$ & $\alpha \beta$ & $(l_{pub}, l_1+1, l_2, \texttt{f}, \texttt{f})$ & $(r_a,r_h)$ \\ \cline{2-5}
		$(l_{pub}, l_1=l_{pub}, l_2 < l_{pub}, \texttt{t}, \texttt{f})$ & Adv. extends $chain_2$ & $\alpha(1-\beta)$ & $(l_{pub}, l_1, l_2+1, \texttt{t}, \texttt{f})$ & $(r_a,r_h)$ \\ \cline{2-5}
		& Type 1 or 2 extends $chain_{pub}$ & $1-\alpha$ & $(l_{pub}+1, l_1+1, l_2, \texttt{t}, \texttt{f})$ & $(r_a,r_h+v_b)$\\ 
		\hline
		
		Case 2.2 & Adv. extends $chain_1$ & $\alpha \beta$ & $(l_{pub}, l_1+1, l_2, \texttt{f}, \texttt{f})$ & $(r_a,r_h)$ \\ \cline{2-5}
		$(l_{pub}, l_1=l_{pub}, l_2 = l_{pub}, \texttt{t}, \texttt{f})$ & Adv. extends $chain_2$ & $\alpha(1-\beta)$ & $(l_{pub}, l_1+1, l_2+1, \texttt{f}, \texttt{t})$ & $(r_a,r_h)$ \\ \cline{2-5}
		& Type 1 or 2 extends $chain_{pub}$ & $1-\alpha$ & $(l_{pub}+1, l_1+1, l_2, \texttt{t}, \texttt{f})$ & $(r_a,r_h+v_b)$\\ 
		\hline
		
		Case 3.1 & Adv. extends $chain_2$ & $\alpha$ & $(l_{pub}, l_1, l_2+1, \texttt{f}, \texttt{f})$ & $(r_a,r_h)$ \\ \cline{2-5}
		$(l_{pub}, l_1 > l_{pub}, l_2 < l_1, \texttt{f}, \texttt{f})$ & Type 2 extends $chain_1$ & $\mu$ & $(l_{pub}, l_1, l_2+1, \texttt{f}, \texttt{f})$ & $(r_a,r_h)$                                                            \\ \cline{2-5}
		& Type 1 extends $chain_{pub}$ & $1 - \alpha - \mu$ & $(l_{pub}+1, l_1, l_2, \texttt{f}, \texttt{f})$ &$(r_a,r_h+v_b)$ \\ 
		\hline
		
		Case 3.2 & Adv. extends $chain_2$ & $\alpha$ & $(l_{pub}, l_1, l_2+1, \texttt{f}, \texttt{f})$ & $(r_a,r_h)$ \\ \cline{2-5}
		$(l_{pub}, l_1 > l_{pub}, l_2 = l_1, \texttt{f}, \texttt{f})$  & Type 2 extends $chain_1$ & $\mu$ & $(l_{pub}, l_1+1, l_2+1, \texttt{f}, \texttt{t})$ & $(r_a,r_h)$ \\ \cline{2-5}
		& Type 1 extends $chain_{pub}$ & $1-\alpha -\mu$ & $(l_{pub}+1, l_1, l_2, \texttt{f}, \texttt{f})$ & $(r_a,r_h+v_b)$ \\ 
		\hline     
		
		Case 3.3 & Adv. extends $chain_2$ & $\alpha$ & $(l_{pub}, l_1, l_2+1, \texttt{f}, \texttt{f})$ & $(r_a,r_h)$ \\ \cline{2-5}
		$(l_{pub}, l_1 = l_{pub}, l_2 < l_1, \texttt{f}, \texttt{f})$ & Type 2 extends $chain_1$ & $\mu$ & $(l_{pub}, l_1, l_2+1, \texttt{f}, \texttt{f})$ & $(r_a,r_h)$                                                            \\ \cline{2-5}
		& Type 1 extends $chain_{pub}$ & $1 - \alpha - \mu$ & $(l_{pub}+1, l_1+1, l_2, \texttt{t}, \texttt{f})$ &$(r_a,r_h+v_b)$ \\ 
		\hline
		
		Case 3.4 & Adv. extends $chain_2$ & $\alpha$ & $(l_{pub}, l_1, l_2+1, \texttt{f}, \texttt{f})$ & $(r_a,r_h)$ \\ \cline{2-5}
		$(l_{pub}, l_1 = l_{pub}, l_2 = l_1, \texttt{f}, \texttt{f})$  & Type 2 extends $chain_1$ & $\mu$ & $(l_{pub}, l_1+1, l_2+1, \texttt{f}, \texttt{t})$ & $(r_a,r_h)$ \\ \cline{2-5}
		& Type 1 extends $chain_{pub}$ & $1-\alpha -\mu$ & $(l_{pub}+1, l_1+1, l_2, \texttt{t}, \texttt{f})$ & $(r_a,r_h+v_b)$ \\ 
		\hline

		
		Case 4 (Double spending) & (instantaneous transition) & $1$ & $(l_{pub}=0, l_1=l_1-l_{pub},$  & $(r_a+(l_2-n_{t2})v_b+v_t,$ \\ 
		S s.t. $l_2 {\geq} l_{pub} {\geq} 6$ & $chain_{pub}=chain_2$& &$l_2=0, s_{(pub,1)}, \texttt{t})$ & $r_h-(l_{pub}-n_{t2})v_b)$\\
	\end{tabular}
    \end{adjustbox}
	\renewcommand{\arraystretch}{1}
	\label{transition}
\end{table}

\begin{table}[htbp]
	\centering
	\scriptsize
	\caption{States and transitions of the DPC attack in the presence of pessimistic type 2 miners.} 
	
	\renewcommand{\arraystretch}{1.2}
	\begin{adjustbox}{angle=270}
		\begin{tabular}{p{4.6cm}|p{4.4cm}|p{1cm}|l|p{2.8cm}}
			
			\textbf{State S} & \textbf{Event} & \textbf{Prob.} & \textbf{Destination state} & \textbf{Revenue} \\ 
			$(l_{pub}, l_1, l_2, {sync}_{(pub,1)}, {sync}_{(1,2)})$ & & & & $(r_a,r_h)$ \\
			\hline
			
			Case 0 (Contains Init state) &  Adv. extends $chain_1 {=} chain_2$ & $\alpha$ & $(l_{pub},l_1+1,l_2+1,\texttt{f},\texttt{t})$ &  $(r_a,r_h)$\\ \cline{2-5}
			$(l_{pub}, l_1, l_2, \texttt{t}, \texttt{t})$ & Type 1 or 2 extends $chain_{pub}$ & $1-\alpha$ & $(l_{pub}+1,l_1+1,l_2,\texttt{t},\texttt{f})$) & $(r_a,r_h+v_b)$ \\ 
			\hline
			
			Case 1 & Adv. extends $chain_2$ & $\alpha$ & $(l_{pub}, l_1, l_2+1, \texttt{f}, \texttt{f})$ & $(r_a,r_h)$ \\ \cline{2-5}
			$(l_{pub}, l_1 > l_{pub}, l_2, \texttt{f}, \texttt{t})$ & Type 1 extends $chain_{pub}$ & $1-\alpha-\mu$ & $(l_{pub}+1, l_1, l_2, \texttt{t}, \texttt{f})$ & $(r_a,r_h+v_b)$\\ \cline{2-5}
			& Type 2 adopts $chain_{pub}$ & $\mu$ & $(l_{pub}, l_1-1, l_2, \texttt{t}, \texttt{f})$ & $(r_a,r_h+v_b)$\\
			\hline
			
			
			Case 2.1 & Adv. extends $chain_1$ & $\alpha \beta$ & $(l_{pub}, l_1+1, l_2, \texttt{f}, \texttt{f})$ & $(r_a,r_h)$ \\ \cline{2-5}
			$(l_{pub}, l_1=l_{pub}, l_2 < l_{pub}, \texttt{t}, \texttt{f})$ & Adv. extends $chain_2$ & $\alpha(1-\beta)$ & $(l_{pub}, l_1, l_2+1, \texttt{t}, \texttt{f})$ & $(r_a,r_h)$ \\ \cline{2-5}
			& Type 1 or 2 extends $chain_{pub}$ & $1-\alpha$ & $(l_{pub}+1, l_1+1, l_2, \texttt{t}, \texttt{f})$ & $(r_a,r_h+v_b)$\\ 
			\hline
			
			Case 2.2 & Adv. extends $chain_1$ & $\alpha \beta$ & $(l_{pub}, l_1+1, l_2, \texttt{f}, \texttt{f})$ & $(r_a,r_h)$ \\ \cline{2-5}
			$(l_{pub}, l_1=l_{pub}, l_2 = l_{pub}, \texttt{t}, \texttt{f})$ & Adv. extends $chain_2$ & $\alpha(1-\beta)$ & $(l_{pub}, l_1, l_2+1, \texttt{t}, \texttt{f})$ & $(r_a,r_h)$ \\ \cline{2-5}
			& Type 1 or 2 extends $chain_{pub}$ & $1-\alpha$ & $(l_{pub}+1, l_1+1, l_2, \texttt{t}, \texttt{f})$ & $(r_a,r_h+v_b)$\\ 
			\hline
			
			Case 2.3 & Adv. extends $chain_1$ & $\alpha \beta$ & $(l_{pub}, l_1+1, l_2, \texttt{f}, \texttt{f})$ & $(r_a,r_h)$ \\ \cline{2-5}
			$(l_{pub}, l_1=l_{pub}, l_2 > l_{pub}, \texttt{t}, \texttt{f})$ & Adv. extends $chain_2$ & $\alpha(1-\beta)$ & $(l_{pub}, l_1, l_2+1, \texttt{t}, \texttt{f})$ & $(r_a,r_h)$ \\ \cline{2-5}
			& Type 1 or 2 extends $chain_{pub}$ & $1-\alpha$ & $(l_{pub}+1, l_1+1, l_2, \texttt{t}, \texttt{f})$ & $(r_a,r_h+v_b)$\\ 
			\hline
			
			Case 3.1 & Adv. extends $chain_2$ & $\alpha$ & $(l_{pub}, l_1, l_2+1, \texttt{f}, \texttt{f})$ & $(r_a,r_h)$ \\ \cline{2-5}
			$(l_{pub}, l_1 > l_{pub}, l_2 < l_1, \texttt{f}, \texttt{f})$ & Type 2 adopts $chain_{pub}$ & $\mu$ & $(l_{pub}, l_1-1, l_2, \texttt{f}, \texttt{f})$ & $(r_a,r_h)$                                                            \\ \cline{2-5}
			& Type 1 extends $chain_{pub}$ & $1 - \alpha - \mu$ & $(l_{pub}+1, l_1, l_2, \texttt{t}, \texttt{f})$ &$(r_a,r_h+v_b)$ \\ 
			\hline
			
			Case 3.2 & Adv. extends $chain_2$ & $\alpha$ & $(l_{pub}, l_1, l_2+1, \texttt{f}, \texttt{f})$ & $(r_a,r_h)$ \\ \cline{2-5}
			$(l_{pub}, l_1 > l_{pub}, l_2 = l_1, \texttt{f}, \texttt{f})$  & Type 2 adopts $chain_{pub}$ & $\mu$ & $(l_{pub}, l_1-1, l_2, \texttt{t}, \texttt{f})$ & $(r_a,r_h)$ \\ \cline{2-5}
			& Type 1 extends $chain_{pub}$ & $1-\alpha -\mu$ & $(l_{pub}+1, l_1, l_2, \texttt{t}, \texttt{f})$ & $(r_a,r_h+v_b)$ \\ 
			\hline     
			
			Case 3.3 & Adv. extends $chain_2$ & $\alpha$ & $(l_{pub}, l_1, l_2+1, \texttt{f}, \texttt{f})$ & $(r_a,r_h)$ \\ \cline{2-5}
			$(l_{pub}, l_1 > l_{pub}, l_2 > l_1, \texttt{f}, \texttt{f})$  & Type 2 extends $chain_1$ & $\mu$ & $(l_{pub}, l_1-1, l_2, \texttt{f}, \texttt{f})$ & $(r_a,r_h)$ \\ \cline{2-5}
			& Type 1 extends $chain_{pub}$ & $1-\alpha -\mu$ & $(l_{pub}+1, l_1, l_2, \texttt{t}, \texttt{f})$ & $(r_a,r_h+v_b)$ \\ 
			\hline     
			

			
			Case 4 (Double spending) & (instantaneous transition) & $1$ & $(l_{pub}=0, l_1=0,l_2=0,\texttt{t}, \texttt{t}$  & $(r_a+(l_2-n_{type2})v_b+v_t,$ \\ 
			S s.t. $l_2 {\geq} l_{pub} {\geq} 6$ & $chain_{pub}=chain_2$& && $r_h-(l_{pub}-n_{type2})v_b)$\\
		\end{tabular}
	\end{adjustbox}
	\renewcommand{\arraystretch}{1}
	\label{transition-type2}
\end{table}

\section{Revenue of a DPC Adversary}
\label{profit}

To evaluate the adversary's revenue when it executes the DPC attack we consider the mining reward $v_b$ per block to be equal to 6.25 BTC (which is the case since May 2021). We do not consider the transaction fees as their impact is negligible and because they vary with time~\cite{bar2022werlman}. We evaluate the adversary's revenue depending on the value $v_t$ of a block of transactions it attempts to double spend plus the mining reward $v_b$ per block. In terms of a block of transactions, the adversary might distribute the value of $v_t$ over the transactions of the attack initialization block since each Bitcoin block (1MB) normally includes from 1,500 to 2,500 transactions.


\begin{figure}[t]
	\centering
	\begin{subfigure}[b]{0.48\textwidth}
		\centering
		\includegraphics[width=\textwidth]{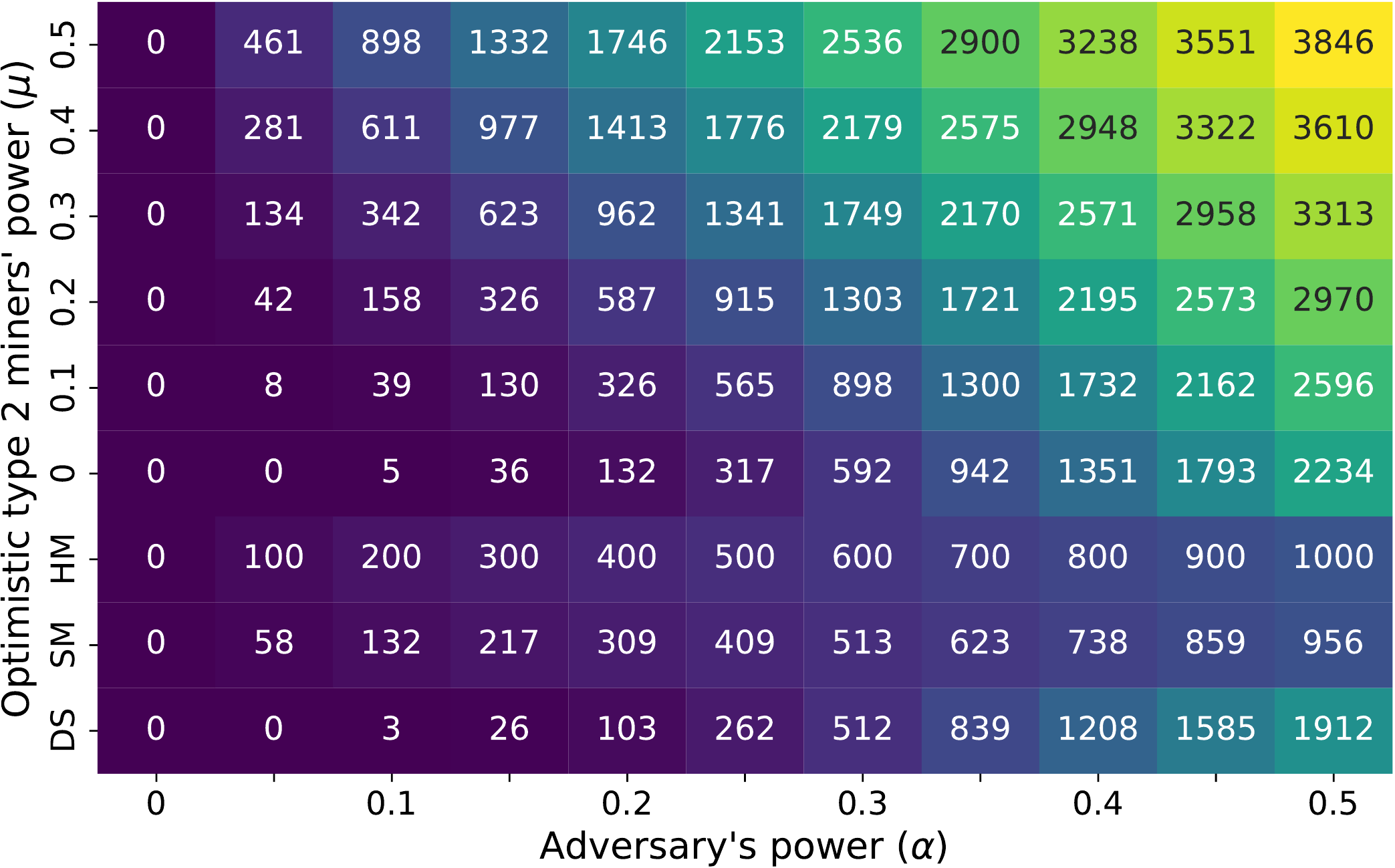}
		\caption{With optimistic type 2 miners.}
	\end{subfigure}
	\hfill
	\begin{subfigure}[b]{0.48\textwidth}
		\centering
		\includegraphics[width=\textwidth]{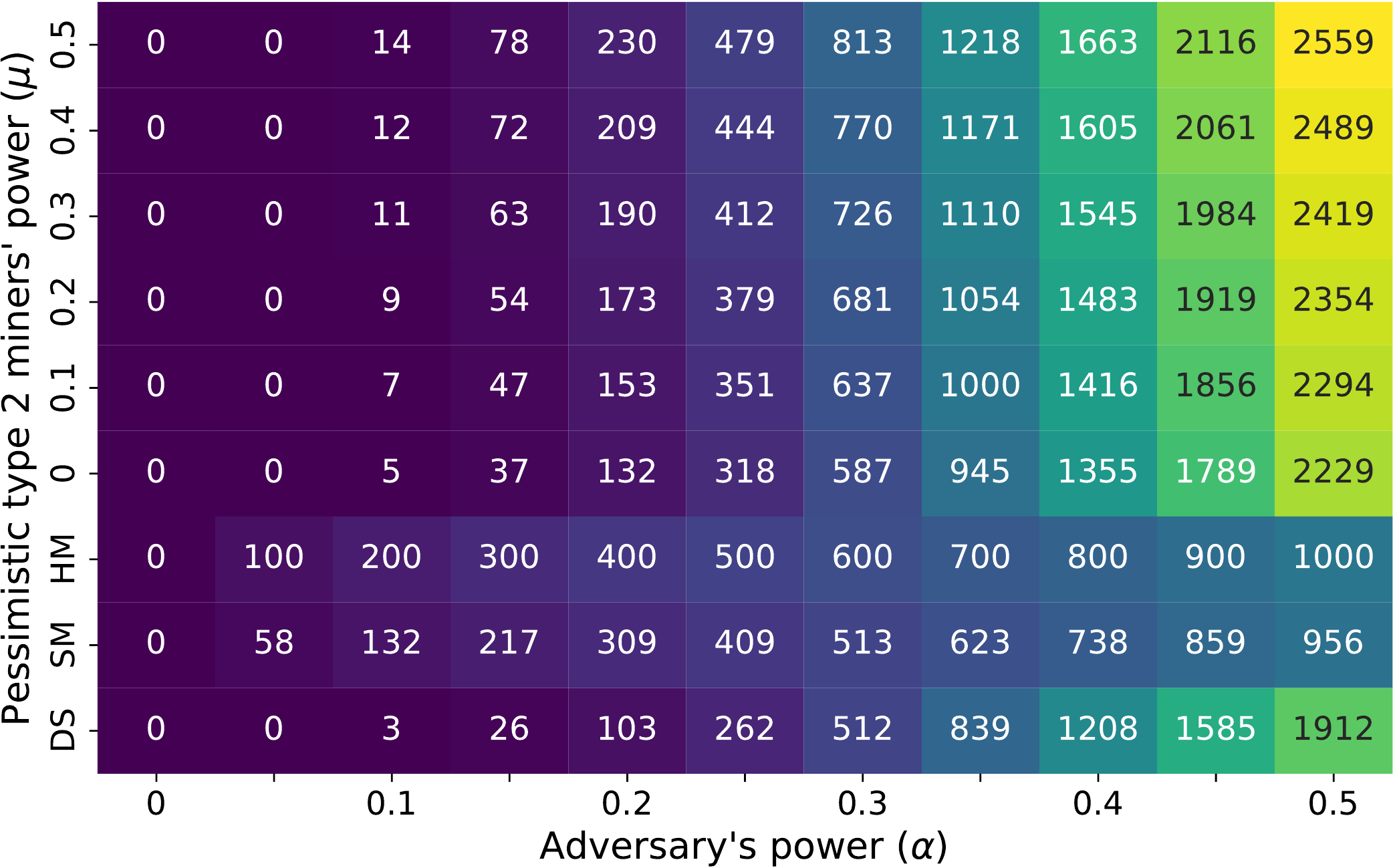}
		\caption{With pessimistic type 2 miners.}
	\end{subfigure}
	
	\caption{Revenue of the adversary when it repetitively attempts to double spend a block with transactions of value $v_t=10v_b$ with the DPC attack and SPV miners, or with a previous mining strategy or attack, over a period of $2,016$ discovered blocks. The bottom 3 rows correspond to previous adversarial mining strategies (classical double spending, selfish mining and honest mining). A lighter color indicates a higher revenue.
    }
	\label{fig:result-10}
\end{figure}

We now show that the DPC attack also increases the revenue a powerful adversary would perceive, assuming that the coin value remains constant despite a successful double spending. 
Upon successfully executing a double spending attack, the on-chain revenue of the adversary consists of the value of the double spent transaction and the mining rewards associated to the blocks that it mined in the longest chain. We consider that  the adversary reinitializes its attack when its double spending chain is too far behind the public chain (cf. Appx.~\ref{sec:reinit}).
We plot the adversary's revenue when $v_t{=}10v_b$, which is reasonable since currently $v_b{=}6.25$~BTC, to compare the different strategies the adversary might follow. 


Fig.~\ref{fig:result-10} compares the expected adversary's revenue for the DPC attack with the revenue it would perceive using honest mining (HM), selfish mining (SM)~\cite{selfish14}, the double spending attack (DS)~\cite{rosenfeld2014analysis}, whose probabilities are represented in the last three lines. We omit the identical energy costs associated to mining for all strategies. 
Selfish mining is less profitable than honest mining, which is expected with a constant mining difficulty~\cite{selfish14,Gervais16}. 
When $v_t=10v_b$ and $\alpha \in [0,0.5]$, the DPC attack is the only strategy for which the adversary's expected revenue can be higher than with honest mining if $\mu > 0$ and $\alpha$ is large enough.    
For example, when $\mu >  0.4$, an adversary with more than $20\%$ of the global hash power (i.e., $\alpha=0.2$) benefits more from the DPC attack than from honest mining. 
Using larger $v_t$ values would also make the DPC attack more profitable than other strategies for given $\mu$ and $\alpha$ values.

\section{Details on Attack Reinitialization}
\label{sec:reinit}

To optimize its revenue and reduce the time necessary to successfully double spend, the adversary can re-initialize its double spending attack. We describe in the following the reinitialization methods it uses with the classical double spending attack and with our DPC attack.

Assuming that transactions require 6 confirmations, with the classical double spending attack the adversary reinitializes its attack if the public chain contains 4 additional blocks compared to the double spending chain. We explain why it is the case in the following. 
Let $l_{ds}$ and $l_{pub}$ be the number of blocks that have been appended to the double spending chain and the public chain since the beginning of the attack. We still assume that the adversary has a fraction $\alpha$ of the global hash power. 
Let $P_{ds}(\alpha)$ be the expected success rate of the double spending attacker based on Nakamoto's evaluation~\cite{nakamoto2008bitcoin}. The classical double spending adversary should reinitialize its attack if $(\frac{\alpha}{1-\alpha})^{l_{pub}-l_{ds}}<P_{ds}(\alpha)$, which comes from the Gambler Ruin model. For $\alpha < 0.5$, when $l_{pub}-l_{ds}< 4$, the inequality $(\frac{\alpha}{1-\alpha})^{l_{pub}-l_{ds}}\geq P_{ds}(\alpha)$ is always verified.

The DPC attack is reinitialized when the adversary estimates that more blocks would need to be generated to successfully double spend from its current state than from the initial state. To provide a fair comparison, we also update traditional double spending attack by using this re-initialization strategy. 
The DPC attacker uses an estimation of the average number of blocks that need to be mined before its attack succeeds for each possible state it might encounter during the attack. In practice, the attacker would run simulations prior to launching its attack to collect these numbers. During the attack, the adversary would then compare the estimated number of necessary blocks until attack success for its current state and for the initial state, and restart its attack if less blocks are required for the attack to succeed starting from the initial state than from its current state.

\end{subappendices}

\end{document}